\documentclass[twocolumn, superscriptaddress, floatfix]{revtex4-2}

\expandafter\let\csname equation*\endcsname\relax
\expandafter\let\csname endequation*\endcsname\relax

\usepackage{graphicx}

\usepackage{algorithm}
\usepackage{algpseudocode}
\usepackage{iopams}
\usepackage{amsmath}
\usepackage{booktabs}
\usepackage{multirow}
\usepackage{amssymb}
\usepackage{bbm}
\usepackage{color}
\usepackage{dcolumn}
\usepackage{bm}
\usepackage{float}
\usepackage[normalem]{ulem}
\usepackage{braket}
\usepackage{changepage}
\usepackage[subrefformat=parens,labelformat=parens,caption=false]{subfig}
\usepackage{enumitem}
\usepackage{bbm}
\usepackage[dvipsnames]{xcolor}
\usepackage{esint}
\usepackage{lineno}
\usepackage{verbatim}
\usepackage{bm}
\usepackage{bbold}
\usepackage{framed}
\usepackage{dsfont}
\usepackage[table]{xcolor}
\usepackage{physics}

\usepackage{tikz}
\usetikzlibrary{quantikz2}
\tikzset{
  evenlayer/.style={dashed, rounded corners, fill=green!20, inner sep=1pt},
  oddlayer/.style={dashed, rounded corners, fill=blue!20, inner sep=1pt},
  layerlabel/.style={label position=below, anchor=north, yshift=-0.2cm}
}
\usepackage{soul}
\usepackage[inner=2.7cm, outer=2cm, bottom=2cm, top=3cm]{geometry}
\usepackage[unicode]{hyperref}
\hypersetup{hidelinks}

\renewcommand\norm[1]{\left\lVert#1\right\rVert}

\newcommand{\XEB}[0]{\mathrm{XEB}}

\begin{document}
\title{Engineered Learning Problems on Noisy Quantum Processors}

\author{Esteban Payares}
\affiliation{IQM France SAS, 4 rue Royale, 75008 Paris, France}

\author{Nadia Milazzo}
\affiliation{IQM France SAS, 4 rue Royale, 75008 Paris, France}

\author{Ghada Ben Slama}
\thanks{Contributions performed during internship at IQM France SAS.}
\affiliation{QbitSoft, 5 Parv. Alan Turing, 75013 Paris, France}

\author{Thomas Cope}
\affiliation{IQM Quantum Computers, Georg-Brauchle-Ring 23-25, 80992 Munich, Germany}
\date{\today}
\begin{abstract}
Despite quantum machine learning being an active topic of research, demonstrations of algorithms run on quantum hardware beating classical methods remain rare. In this work, we demonstrate a significant performance gap using a projected quantum kernel for $n=15,30$ and $45$ qubits over classical methods on an engineered classification problem, and show reproducibility of the results over different quantum processor units. We also demonstrate how error mitigation can widen this gap, both by improving hardware performance but also by increasing the classical difficulty, with different techniques contributing unequally to these phenomena. This work provides strong evidence that superior empirical performance of quantum hardware over all classical techniques is an achievable goal in quantum machine learning, albeit for an artificial problem.
However, as with any application relying on observables, beating state-of-the-art simulation methods remains the primary challenge, as our matrix product state results here show.
\end{abstract}

\maketitle

\section{Introduction}
\label{sec:intro}

Although quantum computers are widely believed to be theoretically more powerful than classical machines, demonstration of this ability on today's devices remains elusive. Algorithms with an unambiguous quantum advantage require advanced, fault tolerant devices, beyond the capability of current technology \cite{Preskill_2018,Gidney2021howtofactorbit,Beverland22}. As such, researchers have turned to more tasks, such as random circuit sampling \cite{Arute_2019,PhysRevLett.127.180501,Morvan2024} to give a more nuanced justification of quantum advantage; with the consequence that even the claim of whether advantage has been achieved is still hotly debated amongst experts \cite{Hangleiter_Advantage}.

Within the sub-field of quantum machine learning, a similar discussion of the usefulness of both current and future devices is ongoing. Well known rigorous speed-ups \cite{liu2021rigorous,muser2024provable,advantageSweke}  remain out of reach of near-term implementations, whilst the merit of approaches more favourable to Noisy Intermediate-Scale Quantum (NISQ) devices, such as variational quantum circuits (VQCs), have met skepticism of their usefulness both from a theoretical perspective \cite{cerezo2024}, as well as their empirical performance \cite{BenchmarkBowles}. Similarly, large scale studies \cite{schnabel2025quantum,BenchmarkBowles} show little to no evidence for improved performance when quantum kernels are employed as a ``one-size-fits-all" approach.  This has led some to question whether quantum machine learning is a sensible field to try to demonstrate quantum advantage at all \cite{QASchuld}. This debate has been further fueled due to the rapid progress in classical machine learning, which is dominated by empirical performance of enormous, resource intensive deep learning models \cite{MLTrends,StanfordHAI2026}. The disconnect between these models and established classical learning theory reduces the impact of theory-driven quantum results.

Rather than trying to demonstrate better performance on end-to-end industrial problems, current cutting edge hardware demonstrations instead aim to establish superior performance of quantum hardware over classical on a task justified simply by the expectation that quantum computers are better suited to the problem, such as random circuit sampling \cite{Arute_2019,PhysRevLett.127.180501,Morvan2024} or simulation of quantum systems \cite{Andersen2025,Shaw2024,IQM_Utility}. One should imagine a scenario where a referee announces a problem to a group of challengers. These challengers then provide solutions to the referee, who judges their quality. If only a challenger with access to a quantum computer can provide a good quality solution, then it is reasonable to claim that quantum supremacy has been achieved \cite{Preskill_2012}.

In practice, the authors of proposed advantage demonstrations act as both challenger and referee, and therefore have to assess the quality of their own calculations, which is not typically possible without access to superior quantum devices, or a more powerful classical machine, which would undermine the claim of supremacy. This is not the only possible approach; the referee could rely on one-way functions, as would be the case with Shor's algorithm, or planted-secret proposals \cite{shepherd2009temporally,Bremner2025}. For methods relying on the linear cross-entropy ($\XEB$) such as \cite{Arute_2019,PhysRevLett.127.180501,Morvan2024,Andersen2025}, or observable estimation \cite{Shaw2024,IQM_Utility}, correctness is only inferred by extrapolating from quantum calculations for which classical simulation is possible, such as smaller system sizes or Clifford circuits.

In this work we implement on quantum hardware a proposal for engineering classification tasks for which NISQ devices could outperform classical models, first introduced in \cite{huang2021power}. The idea (explained in detail in Sec.~\ref{subsec:learning_problem}) is to embed data into a quantum circuit in order to create a quantum kernel, and then retroactively assign labels to the data designed to maximise the difference in predictive performance between the quantum kernel and a specific classical kernel. In a referee-challenger experiment as described above, the referee would perform this labelling and ask the challenger to predict the labels of a set of test data points, given access to a training set of data and the corresponding labels, along with a description of the circuits used. A quantum challenger can then obtain high prediction accuracy by reproducing the kernel and training a support vector machine (SVM), whilst the classical challenger is forced to try and simulate the circuits directly, or train a circuit-agnostic classical model, with the latter expected to perform poorly due to the label construction.

This construction is particularly nice because it only requires the referee to have quantum hardware of a quality similar to the quantum challenger, and provides a way to assess comparative performance to both circuit-agnostic and circuit-simulation classical approaches, in the latter case without needing to justify any assumptions made to simplify simulation, since only predictive accuracy matters.

We implement one family of the kernels introduced in \cite{huang2021power} on IQM hardware, and demonstrate that the quantum challenger outperforms a suite of classical circuit-agnostic challengers, on experiments up to 45 qubits. Unlike the ideal case of \cite{huang2021power}, on hardware we risk engineering the labels to fit the noise, rather than the chosen embedding circuit. To avoid this, we ensure the results are reproducible by running the challenger on a different quantum processing unit (QPU) when possible, whilst for the larger experiments choosing a new transpilation designed to produce a different noise profile. We also show that error mitigation can help to revive an advantage in accuracy degraded by noise, although this is only possible if the mitigation helps to correct bias in the features used to construct the kernel, rather than simply magnitude rescaling.

We do not however claim here to demonstrate quantum advantage, since we can show that for this embedding a kernel using matrix product state approximation reaches performance comparable to the quantum challenger. Based on our results, we see direct circuit-simulation approaches as the biggest obstacles this approach faces, which to solve requires close collaboration with the quantum simulation community.

The rest of the paper is organized as follows: in Sec.~\ref{sec:related_works} we discuss related and relevant literature and how it compares to the results of this work. Sec.~\ref{sec:background} introduces classical and quantum kernels, data encoding and all the necessary concepts and notation from \cite{huang2021power} to setup the experiments presented. In Sec.~\ref{sec:methods} we describe in detail all the ingredients of an experimental run, including the quantum circuits run on hardware, the hyperparameter optimization for the machine learning models and the error mitigation techniques considered. Sec.~\ref{sec:results} presents the results of our experiments and Sec.~\ref{sec:discussion} the main conclusions and future directions.

\section{Related Works}
\label{sec:related_works}

Although the work of Huang et al. \cite{huang2021power} is well known in the field of quantum machine learning, 
quantum kernel experiments implemented on hardware, with the goal of verifying separation between quantum and classical predictive performance, remain limited. This is perhaps due to the artificial nature of the learning problem, since there exist works applying quantum kernels to e.g. fraud detection \cite{Fraud_Kyriienko,AQACYB}, finance \cite{MiyabeFinance}, time series modelling \cite{TimeSeriesPolche}, high energy physics \cite{HEPKernel1,HEPKernel2} and more. Nevertheless, there have been works which have followed-up the label engineering results of \cite{huang2021power}, often in combination with a study of quantum kernel performance for a desired application, and we try to list the most relevant here.

The authors of \cite{Gupta_2025} show a gap between noiseless quantum kernels and a variety of classical models for features taken from earth observation data, whilst \cite{Freq_Kernels} repeat the process for data representing electrical currents.
In \cite{Egginger_Hyperparam} the authors investigate how the choice of total evolution time and the size of the projected subsystem  influence the final prediction separation, albeit for small simulated examples. They found that if the total evolution time is small, then a performance gap is not possible, due to the geometric distance being small. They also observe that increasing the size of the projected subsystem may increase the gap (something we do not consider here). 

The authors of \cite{SakhnenkoArtificial} introduce an alternative way to engineer labels for quantum kernels which does not use any classical kernel, and allows control over the complexity of the resulting function. Although only analysis on noiseless simulation is performed, they are also able to show a performance gap against a classical predictive model. A robustness comparison of the two label-engineering approaches on quantum hardware would be an interesting follow up experiment.

Papers which use hardware to predict engineered labels include \cite{Yin_Photon}, which uses a noiseless simulation to engineer the label for a fidelity-based kernel, with an integrated photonic processor able to outperform classical models, and \cite{Graph_Data}, where the authors use a kernel calculated using a neutral atom quantum processor to relabel the data, but do not appear to investigate the reproducibility of predictions over different hardware/kernel instances (the ``challenger" referred to in our paper). 

Other works study the optimisation of quantum kernel hyperparameters through the lens of kernel performance, such as \cite{schnabel2025quantum,canatar2023bandwidth}. Indirectly related to the work presented here are the results of \cite{slattery2023numerical}, who show that optimising hyperparameters for a given task can lead to easier simulation classically.

Mitigation for the purpose of improving quantum kernel performance was investigated in \cite{hubregtsen2022training,singh2025resource,peters2021machine}. The techniques of \cite{hubregtsen2022training} were designed for the fidelity kernels, and do not translate over to our scenario due to the focus on depolarising noise, which has little impact on our kernels due to the explanation in Sec.~\ref{subsec:compare_ideal}. The readout error mitigation approach specific to \cite{peters2021machine} is also designed for fidelity kernels. Similar to us, the authors of \cite{singh2025resource} use dynamical decoupling and zero noise extrapolation (as well as probabilistic error cancellation) to mitigate their kernels, and use a paired $t$-test to evaluate performance against classical kernels, but focus on fidelity kernels for real-world datasets.

\section{Background}
\label{sec:background}

\subsection{Classical kernels}
\label{subsec:cl-kernel}
Kernel methods are widely used in machine learning to solve a variety of non-linear problems, such as classification and regression problems. They are also used as a theoretical tool for understanding e.g., deep neural networks \cite{NeuralTangent}, an insight which has been extended to quantum variational models \cite{schuld2021supervised}.

They rely on the concept of a kernel function, which computes a similarity measure between data points; this is typically a function of high dimensional transformations of the data, known as \textit{feature vectors}. The inner product of two such vectors defines a kernel $k: \mathcal{X} \times \mathcal{X} \rightarrow \mathbb{K}$:
\begin{equation}
    \kappa(\mathbf{x},\mathbf{x'}) = \langle\phi(\mathbf{x}), \phi(\mathbf{x'})\rangle_{\mathcal{F}}
\end{equation}
where $\mathcal{X}$ is the space of the input data $\{\mathbf{x}_i\}_{i=1}^N$ and $\mathbb{K} = \mathbb{R}$ or $\mathbb{K} = \mathbb{C}$. The mapping of the input data to a feature space $\mathcal{F}$, usually higher-dimensional, makes the problem easier to solve (for example, for a classification problem, different classes can be linearly separated). This mapping is achieved by a feature map $\phi: \mathcal{X} \rightarrow \mathcal{F} $ and the vectors $\phi(\mathbf{x}) \in \mathcal{F}$ are the feature vectors. 
The kernel assigns a similarity measure between the two data points $\mathbf{x}$ and $\mathbf{x'}$ in an alternative feature space $\mathcal{H}_\kappa$ of a kernel, called the reproducing kernel Hilbert space (RKHS). One key result of kernel theory is the representer theorem \cite{scholkopf2001generalized}, which states that the solution $f^*(\mathbf{x}) \in \mathcal{H}_\kappa$ of a machine learning optimisation problem can be represented as an expansion of kernel functions:
\begin{equation}
    f^*(\mathbf{x}) = \sum_{i=1}^N \alpha_i \kappa(\mathbf{x}, \mathbf{x}_i)
    \label{eq:k-expansion}
\end{equation}
for some real coefficients $\alpha_i$. This important result is central to kernel methods and it is often referred to as the kernel trick, since this problem can be solved directly using convex optimisation to find the coefficients $\alpha_i$, provided one can calculate the function $\kappa(\mathbf{x}, \mathbf{x}')$. This avoids the explicit calculation of the feature map, which is typically more expensive to compute.
Mercer's theorem \cite{scholkopf2001learning} guarantees the existence of $\phi$ given that the kernel function $\kappa$ is symmetric and positive semi-definite. These conditions are usually verified on the kernel matrix $K$ with entries:
\begin{equation}
    K_{ij} = \kappa(\mathbf{x}_i,\mathbf{x}_j)
\end{equation}
Choosing the optimal kernel function for a learning problem is not a trivial task; one common choice in classical machine learning literature is the radial basis function (RBF) kernel, also known as Gaussian kernel, $\kappa(\mathbf{x},\mathbf{x'})=\exp(-\gamma\norm{\mathbf{x}-\mathbf{x'}}^2)$, where the parameter $\gamma = \frac{1}{2\sigma^2}$ is the inverse \textit{bandwidth} and needs usually to be fine-tuned to achieve optimal performance. This parameter defines the radius of influence of a single data point: low values of $\gamma$ indicate a broader influence, and they usually provide better generalization of the learning model. Having to tune a single parameter to control locality is one of the properties that make RBF kernels a popular choice. In addition to that, they also have the universal approximation property, that is, they can be used to approximate any continuous function to arbitrary precision. The RBF kernel, in combination with Support Vector Machines, will be used in our work to benchmark the prediction performance of quantum kernels, introduced in the next section.

\subsection{Quantum kernels} 
Quantum kernel methods implement the feature map $\phi(\mathbf{x}_i)$ described in Sec.~\ref{subsec:cl-kernel} through a data-encoding quantum circuit. In the original introduction of quantum kernels \cite{havlivcek2019supervised, schuld2019quantum} the inner product used to define the kernel was $\kappa(\mathbf{x}, \mathbf{x'})=\vert \langle \phi(\mathbf{x}),\phi(\mathbf{x'})\rangle \vert^2$, called the \emph{fidelity kernel}, more easily recognisable as the inner product of Hermitian operators $\kappa(\mathbf{x}, \mathbf{x'})=\mathrm{Tr}[\rho(\mathbf{x}),\rho(\mathbf{x'})]$. For sufficiently expressive choices of $\phi(\mathbf{x}_i)$, \cite{huang2021power} noted that such overlaps become exponentially small, making the kernel matrix effectively the identity, and unable to generalise. The learning theory of this issue was expanded in \cite{kubler2021inductive}, whilst the conditions for exponential concentration were studied in \cite{ThanasilpConcentration}.
To counter this, \cite{huang2021power} also introduced the idea of \emph{projected quantum kernels}, in which kernel elements are defined using a positive semi-definite function applied to a vector of local observables; for example, 
\begin{equation}
    K_{ij}=\exp\left(-\gamma \sum_{k} \lVert\rho_k(\mathbf{x}_i)-\rho_k(\mathbf{x}_j)\rVert_F^2\right)
    \label{PQK}
\end{equation} 
where $\rho_k(\mathbf{x}_i)=\mathrm{Tr}_{j\neq k}(\rho(\mathbf{x}_i))$ is the reduced density matrix of qubit $k$. Since a single-qubit state is fully determined by its three Pauli expectations, we work directly with the projected features:
\begin{equation}
f_{k,P}(\mathbf{x}) = \operatorname{Tr}\!\big[\rho(\mathbf{x})\,P_k\big], \quad P\in\{X,Y,Z\},\ k=1,\dots,n,
\end{equation}
where $P_k = I^{\otimes(k-1)}\otimes P\otimes I^{\otimes(n-k)}$ denotes the Pauli operator $P$ acting on qubit $k$. 

Thus, Eq.~\eqref{PQK} becomes an RBF kernel on $f(\mathbf{x})\in\mathbb{R}^{3n}$, $K_{ij}=\exp(-\gamma_{\mathrm{eff}}\lVert f(\mathbf{x}_i)-f(\mathbf{x}_j)\rVert^2)$. We follow \cite{huang2021power} and use a variance-normalized bandwidth $\gamma_{\mathrm{eff}}$:
\begin{equation}
\gamma_{\mathrm{eff}} = \frac{\gamma_0}{n_f\,\sigma_f^2},
\label{eq:bandwidth}
\end{equation}
where $\sigma_f^2$ is the (pooled) variance of the projected features over the training set and $n_f$ the number of features.

Due to its linear scaling with the size of the data and the vanishing overlap issues arising for fidelity kernels, projected kernels have been seen as the more practical approach, being deployed in a range of experiments \cite{MiyabeFinance,UtroToxicity,TimeSeriesPolche}; we will follow these results and also use a kernel of the form in Eq.~\eqref{PQK} in our experiments. Not all PQK embeddings are practical, though; the results of \cite{ThanasilpConcentration,BPEquivalence} show that for embeddings where $\rho$ follows a volume law, quantum kernels can collapse to the constant kernel.

\subsection{Engineered learning problem}
\label{subsec:learning_problem}
The idea of constructing labels artificially to achieve high performance of a desired quantum kernel was already present in the early literature of quantum kernels \cite{havlivcek2019supervised, schuld2019quantum}. The artificial labels construction introduced later in \cite{huang2021power} was designed to provide a practical and efficient way to demonstrate an empirical advantage using quantum machine learning,  aiming to separate an explicit quantum and classical kernel model, rather than via theoretical learning complexity assumptions. More precisely, this separation can be quantified with the complexity of the model, the quantity that determines the prediction error bound (in the ideal case - see Sec.~\ref{subsec:number_of_data} for more details on the noisy case). Smaller values of the complexity correspond to better generalization of the model on unseen data. If $s_{K_C}$ is the complexity of the classical model defined by the classical kernel $K_C$ and $s_{K_Q}$ is that of the quantum model, defined by the quantum kernel $K_Q$, then the potential advantage of the quantum model to predict better than the classical one depends on the largest possible separation between $s_{K_Q}$ and $s_{K_C}$. This is what the protocol presented in \cite{huang2021power} implements: for a dataset $\{(x_i,y_i)\}_{i=1}^N$, it aims at maximizing the classical model complexity, defined as $s_{K_C} = y^T K_C^{-1} y $, 
when the quantum complexity $s_{K_Q}$ is set to be $1$. This optimization problem for the labels $\mathbf{y}$ can be formulated as a generalized eigenvalue problem, whose solution has eigenvalue:
\begin{equation}
    g^2 = \norm{\sqrt{K_Q}\sqrt{K_C}(K_C + \lambda I)^{-2}\sqrt{K_C}\sqrt{K_Q}}_{\infty}
    \label{ggen}
\end{equation}
The quantity $g$ in Eq.~\eqref{ggen} is called the geometric difference and $\lambda > 0$ is a regularization parameter. 
More details on the construction of the engineered dataset are presented in Sec.~\ref{subsec:hyperparams}.

\subsection{Hamiltonian evolution encoding}
\label{subsec:encoding}

The data-encoding circuit used in this work is a Trotterized time evolution of the $1$-dimensional inhomogeneous XXX Heisenberg Hamiltonian with nearest-neighbour coupling and no external field:
\begin{equation}
    H = \sum_{i=1}^{n-1} J_iH_i = \sum_{i=1}^{n-1} J_i(X_iX_{i+1}+Y_iY_{i+1}+Z_iZ_{i+1})
\end{equation}
where $X_i,Y_i,Z_i$ are the Pauli operators, $J_i>0$ and $n$ is the number of qubits. This model has been extensively studied in the literature as a platform to understand the underlying physics of quantum many-body systems, and quantum computers have been already used for this purpose \cite{chowdhury2024enhancing, lee2026digital}. Moreover, it has also been used to demonstrate optimizations and improvements for the digitalization of quantum time evolution \cite{yang2026quantum, aktar2026structure} through Trotterization. 

The Trotter decomposition approximates the time evolution operator of the full Hamiltonian $e^{-itH}$ by a series of small evolutions under individuals terms of the Hamiltonian, which are easier to implement experimentally. We choose here a second-order Trotter decomposition \cite{lee2026digital}, which in this case can be expressed in terms of non-commuting even and odd terms as:

\begin{equation}
    e^{-itH} \approx \left(e^{-iH_{even}\frac{t}{2T}}e^{-iH_{odd}\frac{t}{T}}e^{-iH_{even}\frac{t}{2T}}\right) ^T
\end{equation}

where we divide the total evolution time $t$ into $T$ Trotter steps. The data $\{x_k\}$ are encoded as the couplings of the Hamiltonian:

\noindent\resizebox{0.95\columnwidth}{!}{%
 \begin{minipage}{\linewidth}
\begin{align}
   \ket{x_k} = &\left(\prod_{\substack{i=0 \\ i\ \mathrm{even}}}^{n-2} e^{-i \frac{x_{ki}t}{2T} H_i} \prod_{\substack{i=0 \\ i\ \mathrm{odd}}}^{n-2} e^{-i \frac{x_{ki}t}{T} H_i} \prod_{\substack{i=0 \\ i\ \mathrm{even}}}^{n-2} e^{-i \frac{x_{ki}t}{2T} H_i} \right) ^T \nonumber \\ 
   &\bigotimes_{i=1}^{n} \ket{\psi_i}
\end{align}
\end{minipage}%
}

where $\ket{\psi_i}$ is a Haar random single-qubit state fixed for all inputs. The resulting quantum circuit is reported in Sec.~\ref{subsec:circuits}. Time dynamics of $1$-dimensional models are still within reach of classical simulation, especially for short evolution times \cite{osborne2006,vidal2007,haegeman2011,peng2022,wild2023}, while $2$-dimensional problems become more challenging for large enough lattice sizes \cite{paeckel2019,verresen2018}. Although the goal of the present work is not to choose classically difficult dynamics, it will be a necessary step for any supremacy experiments.

\section{Methods}
\label{sec:methods}

\subsection{Hardware description}
\label{subsec:qpus}
\begin{figure*}[t]
    \centering
    \subfloat[\texttt{Emerald}\label{fig:emerald_chain_e}]{%
      \includegraphics[width=0.49\textwidth]{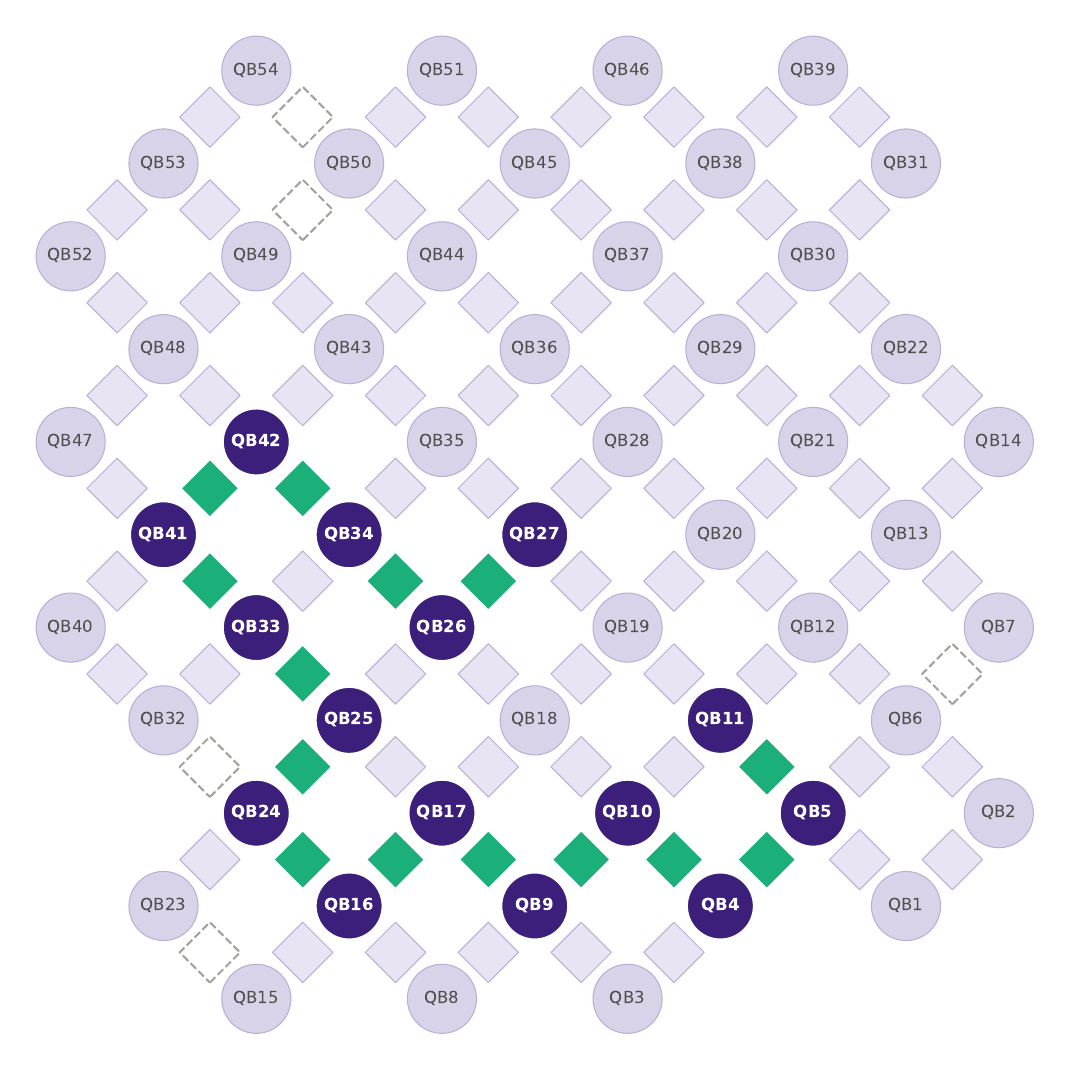}}\hspace{2em}%
    \subfloat[\texttt{Garnet}\label{fig:emerald_chain_g}]{%
      \includegraphics[width=0.31\textwidth]{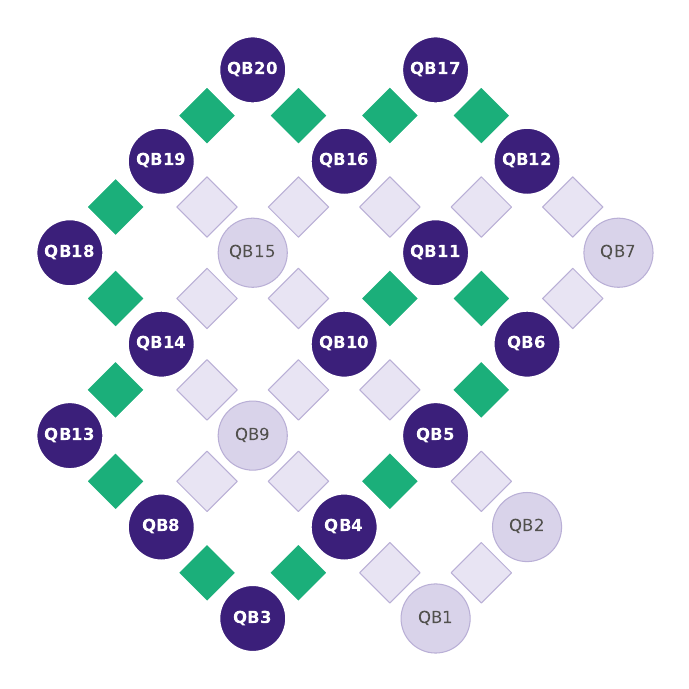}}
    \caption{Qubits (circles) and tunable couplers (diamonds) of IQM \texttt{Emerald} and \texttt{Garnet} on their square lattices; dashed diamonds are unavailable couplers. Highlighted: the $15$-qubit chains of the referee (\texttt{Emerald}) and challenger (\texttt{Garnet}) kernels in the $T{=}10$ error mitigation experiments (Sec.~\ref{subsec:em_methods}). Running the challenger on a different device from the referee tests cross-device reproducibility and ensures that a gap in accuracy between quantum and classical predictions is not caused by fitting labels to the noise of one device.}
    \label{fig:emerald_chain}
\end{figure*}

For the results presented in this work we used two IQM systems, IQM \texttt{Emerald} and IQM \texttt{Garnet}, two superconducting qubit chips with 54 and 20 qubits respectively. The qubits are arranged in a square lattice (Fig.~\ref{fig:emerald_chain}) and connected by tunable couplers (more details about the coupling scheme can be found in \cite{marxer2023long}). The data encoding chosen to implement the feature map of the quantum kernels is optimally implemented on the QPUs by choosing qubits on a $1$D chain, given that the Heisenberg Hamiltonian only includes nearest-neighbour interactions. We use the IQM Qubit Selector \cite{iqm_qubit_selector} to select the best layout on the QPU based on the calibration data at the time of the experiments. More details on the resulting transpiled depth of the data-encoding quantum circuits are given in Sec.~\ref{subsec:circuits}.

Central to the design of our experiments is their reproducibility on different quantum hardware; this is important to ensure that our results can be verified at a later time and that they are independent of the specific noise profile characterising a QPU. For experiments with a number of qubits smaller than $20$, we can easily test reproducibility using the two different chips described above. When the number of qubits increases, we need to engineer chains on the same chip such that they share the smallest number of qubits and couplers possible, while keeping a low cost. More precisely, for each experiment we will need two chains, to calculate the referee and the challenger quantum kernels. We can fix the first layout as before using the IQM Qubit Selector, while the second is the result of a trade-off between shared couplers and cost. Among the feasible $n$-qubit chains supported by the chip, we filter for the 20 with minimal coupler overlap relative to the first chain, ultimately selecting the candidate with the lowest cost under the Qubit Selector cost function (the resulting pairs are shown in Figs.~\ref{fig:chains_30q} and~\ref{fig:chains_45q}).

\subsection{Circuits description}
\label{subsec:circuits}
\begin{figure}[t]
\centering
\resizebox{\linewidth}{!}{%
\begin{quantikz}
\setwiretype{n} & & & & & & & & &\\
\lstick{$q_0$} & \gate[2]{U_0(\frac{\theta_{k_{0}}}{2})} \gategroup[6,steps=3,style={draw=black,thick,rounded corners,inner sep=3pt}]{} & & \gate[2]{U_0(\frac{\theta_{k_{0}}}{2})}\gategroup[6,steps=2,style={dashed,rounded corners,fill=blue!20, inner sep = 1pt}, background, label style={label
position=below,anchor=north,yshift=-0.2cm}]{} & \gate[2]{U_0(\frac{\theta_{k_{0}}}{2})} \gategroup[6,steps=3,style={draw=black,thick,rounded corners,inner sep=3pt}]{} & & \gate[2]{U_0(\frac{\theta_{k_{0}}}{2})} &\\
\lstick{$q_1$} & & \gate[2]{U_1(\theta_{k_{1}})} & & & \gate[2]{U_1(\theta_{k_{1}})} &\\
\lstick{$q_2$} & \gate[2]{U_2(\frac{\theta_{k_{2}}}{2})} & & \gate[2]{U_2(\frac{\theta_{k_{2}}}{2})} & \gate[2]{U_2(\frac{\theta_{k_{2}}}{2})} & & \gate[2]{U_2(\frac{\theta_{k_{2}}}{2})} &\\
\lstick{$q_3$} & & \gate[2]{U_3(\theta_{k_{3}})} & & & \gate[2]{U_3(\theta_{k_{3}})} &\\
\lstick{$q_4$} & \gate[2]{U_4(\frac{\theta_{k_{4}}}{2})} & & \gate[2]{U_4(\frac{\theta_{k_{4}}}{2})} & \gate[2]{U_4(\frac{\theta_{k_{4}}}{2})} & & \gate[2]{U_4(\frac{\theta_{k_{4}}}{2})} &\\
\lstick{$q_5$} & & \vdots & & & \vdots & \vdots & \\
\setwiretype{n} & \vdots &  & \vdots & \vdots & & \vdots & 
\end{quantikz}
}
\caption{Two Trotter steps (black boxes) of the second-order Trotterization of the Heisenberg Hamiltonian with open boundary conditions, encoding the data point $\mathbf{x}_k$. The two adjacent even layers in the coloured box are merged into a single layer $U_{even}(\boldsymbol{\theta}_k)$, so that $T$ steps require $2T+1$ layers.}
\label{fig:trotter-circ}
\end{figure}
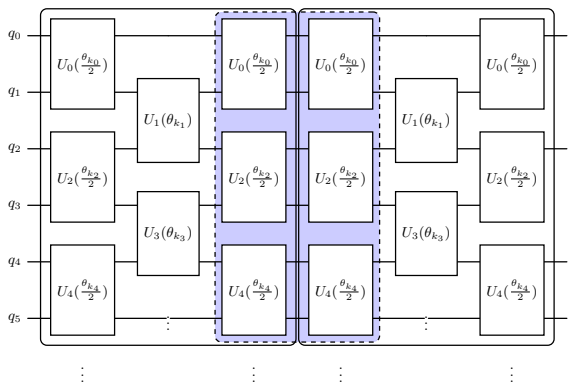

We implement the optimised second-order Trotter decomposition presented in \cite{chowdhury2024enhancing} with open boundary conditions. We recall here the basic building block for the time evolution seen in Sec.~\ref{subsec:encoding}:
\begin{equation}
    U_j(\theta_{k_j}) = e^{-i \frac{\theta_{k_j}}{2}(X_jX_{j+1}+Y_jY_{j+1}+Z_jZ_{j+1})}
\end{equation}
where $\theta_{k_j} = 2\mathbf{x}_{k_j}t/T$ for bond $j$, with the qubits labelled $0,\dots,n-1$ along the chain. One Trotter step can be defined in terms of even and odd layers:
  \begin{equation}
      U_{even}(\boldsymbol{\theta}_k) = \prod_{\substack{j=0 \\ j\ \mathrm{even}}}^{n-2} U_j(\theta_{k_j}), \quad U_{odd}(\boldsymbol{\theta}_k) =
  \prod_{\substack{j=1 \\ j\ \mathrm{odd}}}^{n-2} U_j(\theta_{k_j})
  \end{equation}
The second order Trotter approximation can be written in terms of even and odd layers as:
\begin{equation}
      e^{-itH/T} \approx U_{even}(\boldsymbol{\theta}_k/2)U_{odd}(\boldsymbol{\theta}_k)U_{even}(\boldsymbol{\theta}_k/2)
  \end{equation}
The resulting circuit is shown in Fig.~\ref{fig:trotter-circ}.

Merging the last even layer of each Trotter step with the first even layer of the following step, the second-order Trotter decomposition with $T$ steps can be achieved with $2T+1$ layers (see \cite{chowdhury2024enhancing} for more details on the implementation of the $U_j$ blocks). This decomposition is advantageous as the depth of the quantum circuits does not increase when we add qubits to the encoding, thus allowing us to easily scale the number of qubits in our experiments. The depths of the quantum circuits after transpilation at varying number of Trotter steps are reported in Table~\ref{tab:depths}, together with the number of 2-qubit gates, which does change with the number of qubits.

\begin{table}[t]
    \centering
    \begin{tabular}{c|c|c}
    \toprule
    Trotter step & Circuit depth & No. of 2-q gates \\
    \midrule
    3 & 45 & 147 \\
    4 & 57 & 189 \\
    5 & 69 & 231 \\
    6 & 81 & 273 \\
    10 & 129 & 441 \\
    \bottomrule
    \end{tabular}
    \caption{We report the depth of the data-encoding quantum circuit for $15$ qubits at different number of Trotter steps. Note that the depth will not change if we increase the number of qubits, thanks to the optimised second-order Trotter decomposition. The number of 2-qubit gates will still increase (here the 2-q gates are CZ gates, native to IQM chips.) }
    \label{tab:depths}
\end{table}

\textbf{Encoding hyperparameters}: we fix the total evolution time to $t = n/3$ and vary the number of Trotter steps from $T = 3$ to $T = 10$ across system sizes of $n \in \{15, 30, 45\}$ qubits. Each of the $n - 1$ bonds along the chain encodes a single input feature via the rotation angle $\theta_{k_j} = 2\mathbf{x}_{k_j} t / T$, so that the features directly dictate the rotation angles. In our experiments, the initial product state $\bigotimes_i \ket{\psi_i}$ is prepared by a wall of single-qubit rotations $R_z(\alpha_i)R_y(\beta_i)R_x(\delta_i)$, with randomly drawn angles, fixed for all inputs.

We use the Fashion-MNIST dataset restricted to classes $0$ (T-shirt/top) and $3$ (dress), following the TensorFlow Quantum implementation of~\cite{huang2021power}. This relabelling framework has also been extended to various other datasets~\cite{Gupta_2025}. Pixel values are scaled to $[0,1]$, and each image is projected onto its leading $n - 1$ principal components, assigned one per nearest-neighbour bond along the chain.

\subsection{Engineered labels and model selection}\label{subsec:hyperparams}
We follow the labelling construction of~\cite{huang2021power}. Synthetic labels are generated as $y = \sqrt{K_Q}\,v$, where $v$ is the eigenvector with the largest eigenvalue of the matrix in Eq.~\eqref{ggen}, thresholded at their median and then flipped with probability $0.05$. The classical kernel and the regularization parameter $\lambda$ represent the strongest classical adversary, defined as the configuration that minimizes the geometric difference $g$ subject to the training-error constraint $g_{\mathrm{tra}} < 0.045$ from~\cite{huang2021power}. This search is conducted over RBF kernels with bandwidths $\gamma_C \in \{2^{-2}, 2^{-1}, \dots, 2^{6}\}/(d\sigma_x^2)$ and $100$ logarithmically spaced values $\lambda \in [10^{-9}, 1]$ (Table~\ref{tab:hyperparams}).

While Huang et al.\ tune the quantum bandwidth for prediction, they do not specify how to select it for the label-generating kernel; we determine this parameter via a maximin criterion,
\begin{equation}
    \gamma_Q^* = \arg\max_{\gamma_Q} \min_{\gamma_C} g(\gamma_Q, \gamma_C),
\end{equation}
where $\gamma_Q$ is evaluated on the prediction grid ($\{0.25, 0.5, 1, 2, 3, 4, 5, 20\}$ multiples of Eq.~\eqref{eq:bandwidth}) and the minimization is taken over the feasible classical kernels as defined above. The resulting separation $g$ therefore holds against all considered classical bandwidths on the grid, rather than being an artifact of a single favorable classical baseline.

To predict the labels, each model (both quantum and classical) selects its hyperparameters via 5-fold cross-validation on the $100$ training points and is evaluated once on the $50$ held-out test points. For the PQK, the tuned hyperparameters are the kernel bandwidth, swept across the same grid of multiples of Eq.~\eqref{eq:bandwidth}, and the SVM regularization parameter $C$ ($18$ values between $0.006$ and  $1024$); the classical baselines are detailed in Appendix~\ref{app:adversaries}.

\subsection{Error mitigation description}\label{subsec:em_methods}
It is well known that one of the biggest challenges of current quantum computers is noise; this has motivated entire fields dedicated to error mitigation and suppression to (partially) eliminate the effects of noise on results of quantum experiments. We are interested in understanding which techniques are helpful when running quantum kernels on quantum chips, and if they can improve the performance on the learning problem described in Sec.~\ref{subsec:learning_problem}. To this end, we consider several well-known methods for NISQ computation, but also more advanced ``in-house" techniques, which we now briefly present.

\textbf{Zero Noise Extrapolation (ZNE)}: First introduced in \cite{li2017efficient, temme2017error}, the goal is to obtain the noiseless expectation value of an observable extrapolating from expectation values at different noise levels. Initially, one intentionally scales the noise level in the quantum circuit: this can be done in different ways, both at the gate and at the pulse level. 

In this work, we use global unitary folding \cite{giurgica2020digital} which performs a mapping from the unitary circuit $U$ to $U(U^{\dagger}U)^n$, where $n$ is a positive integer; since $U^{\dagger}U=I$, in the ideal case there is no logical effect, but in a noisy circuit the noise increases due to the addition of more gates. We can write the expectation value of an observable $O$ at a rescaled noise level $\lambda'= s\lambda$ as $\expval{O(\lambda')} = \Tr(\rho(s\lambda)O)$, where $\rho(s\lambda)$ is the quantum state prepared by the noise rescaled quantum circuit (we use here $s = [1,3,5]$) and $\lambda$ is the parameter characterising the noise level in the original circuit. Fitting these points (typically polynomial or exponential) allows us to extrapolate the zero noise limit $\expval{O(\lambda = 0)}$. 

We use here the widely-used Richardson extrapolation \cite{temme2017error} due to its simplicity, although there exist more complex approaches aiming to combat overfitting and shot-noise amplification \cite{mohammadipour2025direct}.

\textbf{Readout Error Mitigation (REM)}: based on classical post-processing \cite{maciejewski2020mitigation}, this technique aims to reduce the errors occurring during the quantum circuit measurement. The key idea is that, assuming perfect preparation of basis states, one can retrieve the ideal probability distribution of bitstrings by inverting the confusion matrix $\Lambda$ in $\vec{p}_{noisy} = \Lambda \vec{p}_{ideal}$. The entries of $\Lambda$ are the probabilities of measuring the bitstring $b$ when the ideal bitstring is $b'$, a conversion caused by the readout error process.
Since the full calculation of $\Lambda$ would require exponential resources, one typically makes assumptions regarding its possible structure. We assume correlation errors are negligible with respect to uncorrelated errors on individual qubits, and thus construct $\Lambda$ as the tensor product of $n$ $2\times 2$ matrices $\Lambda_k$, which we calibrate with the \textit{mthree} package \cite{nation2021scalable}.

As the projected quantum kernel only needs single-qubit expectation values, under this assumption each one can be corrected independently, by applying the \textit{mthree} correction to the marginal distribution of its own qubit. This correction can at most undo the uncorrelated, single-qubit readout error, so it can raise the measured expectation values by at most a factor $1/(1-e_0-e_1)$, where $e_0$ ($e_1$) is the probability of reading a qubit as 1 (0) when it was prepared in 0 (1); for our runs this is, on average over the qubits, $2.1\%$ on \texttt{Emerald} and $3.6\%$ on \texttt{Garnet}.

\textbf{Dynamical Decoupling (DD)}: tailored to reduce the effects of system-environment interactions (quantum decoherence), this error suppression technique drives and ``refocuses" idle qubits' states with a series of control pulses  \cite{zhang2014protected}. DD sequences typically consist of a series of $\pi$-pulses that effectively cancel out, e.g., accumulated phase errors. This technique is especially useful when several qubits stay idle during the quantum computation. We include DD in all our experiments and use the default sequences available via IQM's cloud API: the XX sequence, also known as CPMG sequence, the XY4 sequence \cite{pokharel2018demonstration} and a longer sequence XYXYYXYX, used when the idle time far exceeds the duration of a single-qubit gate. Note that inserting any of these sequences corresponds to an identity operation, thus preserving the original action of the circuit. Which sequence is chosen depends on the idle time it aims to fill.

\textbf{Adaptive Pauli Twirling}: 
We also include in this work results using a new error suppression technique, which we refer to as adaptive Pauli twirling (APT). It improves on standard Pauli twirling by exploiting information that the standard procedure discards \footnote{Details to be released in an upcoming publication}.

\subsection{Evaluation of Mitigation and Performance}\label{subsec:compare_ideal}

To assess the effects of error mitigation we take an ideal baseline to compare with every hardware kernel: the ideal features are obtained with statevector simulations of the circuits defined in Sec.~\ref{subsec:circuits}, for the same dataset of size $N{=}600$.
 
Differences from this reference therefore come only from hardware noise and finite sampling, not from the Trotter approximation.

\textbf{Feature decomposition}: We can differentiate the part of the noise that affects the kernel, from the part that does not, by representing the measured features as a shrunk copy of the ideal ones plus a residual;
\begin{equation}
    f_{\mathrm{hw}}(\mathbf{x}) = c\,f(\mathbf{x}) + r(\mathbf{x}),
    \label{eq:split}
\end{equation}
where $c$ is the global scaling factor that provides the best least-squares fit between the hardware features and the ideal ones,
\begin{equation}
    c = \frac{\sum_{\mathbf{x}} \langle f_{\mathrm{hw}}(\mathbf{x}),\, f(\mathbf{x})\rangle}{\sum_{\mathbf{x}} \lVert f(\mathbf{x})\rVert^2},
    \label{eq:c}
\end{equation}
with the sums taken over all 600 points in the pool. The \emph{scaling factor} $c$ measures the fraction of the signal preserved after noise ($c=1$ corresponds to a fully preserved signal) while the \emph{normalized residual} quantifies the part of the error that a single shrinkage factor cannot explain, expressed as a fraction of the ideal signal. 
\begin{equation}
  \lVert r\rVert = \left(\frac{\sum_{\mathbf{x}} \lVert f_{\mathrm{hw}}(\mathbf{x}) - c\,f(\mathbf{x})\rVert^2}{\sum_{\mathbf{x}} \lVert f(\mathbf{x})\rVert^2}\right)^{1/2}
  \label{eq:r}
\end{equation}
Finally, we define the \emph{residual relative to the surviving signal},
\begin{equation}
  \varepsilon = \frac{\lVert r\rVert}{c}.
  \label{eq:effnoise}
\end{equation}

\textbf{Kernel similarity}: To compare kernels, we use the centred kernel alignment (CKA) \cite{JMLR:v13:cortes12a}, defined as the cosine similarity between centred kernel matrices (Eq.~\eqref{Eq:CKA}, Appendix~\ref{app:em_geometry}). We select CKA as metric because support vector machines with an explicit bias term and tuned regularization parameter are invariant to constant offsets and uniform scaling of the kernel matrix, both of which CKA factors out. For completeness, we also report the relative Frobenius error, $d_F(K, K') = \lVert K - K' \rVert_F / \lVert K' \rVert_F$, which penalizes differences in scale and offset as error.

\textbf{Labels and averaging}: Besides the labels generated by each referee kernel (Sec.~\ref{subsec:learning_problem}), we also construct labels using the ideal kernel via the same protocol to score each hardware kernel. While a kernel can exploit its intrinsic noise on self-derived labels, performing well on ideal labels demands capturing the true noiseless structure. All accuracies reflect averages over $10$ subsets of $150$ points ($100$ training, $50$ test) drawn from the $600$-point pool, evaluated across $10$ label assignments per subset. Each mitigated kernel is paired with an unmitigated baseline executed alongside it; for REM, both kernels are computed from the same measurement counts, and we refer to the unmitigated one as the raw-counts run; the APT kernels, run later on the same qubits, are also compared with the raw-counts run. Unless stated otherwise, all accuracies and gaps in this work follow this averaging and these pairings.

\textbf{Corrected resampled $t$-test}: the central quantity of our results is a performance gap between different learning models. This gap is calculated from pairs of average predictive accuracies of two different models, where the average for each model is taken over 10 different sets of engineered labels. 
Additionally, we repeat this calculation several times for different sub-samples of the data; these sub-samples are not independent, as we use resampling from a $600$-point pool to get $10$ different datasets of size $150$. We therefore obtain $10$ potentially correlated pairs that we need to compare to see whether a gap exists, that is, to understand whether the two learning models perform differently on the same task. One standard approach used in machine learning literature to compare performance of different models is the paired $t$-test, whose null hypothesis $H_0$ is that the mean difference between pairs is zero. However, this test assumes independent pairs and is thus not directly applicable in our case. 

A correction to the test for correlated data was proposed in \cite{nadeau1999inference}, where the authors introduce a correction term to mitigate Type I errors (false positives), which occur when one rejects the null hypothesis although it is true. The statistic of the corrected resampled $t$-test is defined as:
\begin{equation}
    t = \dfrac{\frac{1}{M}\sum_{i=1}^M A_i^q - A_i^{cl}}{\sqrt{\left(\frac{1}{M}+\frac{N_\mathrm{test}}{N_{\mathrm{train}}}\right)\sigma^2}}
\end{equation}
where M is the number of pairs, $A_i^q$ and $A_i^{cl}$ are the test accuracies of the quantum and classical models respectively, $\sigma^2$ is the variance of the $M$ pairs and the ratio $N_{\mathrm{test}}/N_{\mathrm{train}}$ is the correction term with respect to a standard $t$-test.
Although the corrected $t$-test only offers an estimate of the true correlation, this estimate is designed to be an overestimate to further discourage Type I errors. We therefore see it as an improvement on the standard uncorrected approach. 
In all the following results, we will say that ``there is a gap" between two models performances when the test rejects $H_0$, while ``no gap" means that we fail to reject $H_0$.

\section{Results}
\label{sec:results}
We first investigate how the gap between quantum and classical models behaves in the presence of noise as the circuit depth and the size of the system grow (Sec.~\ref{subsubsec:unmit_results}), and then apply error mitigation in the regime where the gap is lost (Sec.~\ref{subsubsec:em_results}). We then compare the hardware challenger with a classically simulated matrix product state kernel (Sec.~\ref{subsec:mps}), extend the label construction to new test points (Sec.~\ref{subsec:newlabels}), and study the effect of the dataset size (Sec.~\ref{subsec:number_of_data}).

\subsection{Limitations of Unmitigated Circuits}\label{subsubsec:unmit_results}

To assess whether the theoretical advantage of the ideal kernel persists in the presence of hardware noise at increasing circuit depths, we evaluate performance across Trotter steps $T \in \{3, 4, 5, 6, 10\}$ for $n = 15, 30,$ and $45$ qubits. Adopting the protocol established in Secs.~\ref{subsec:hyperparams} and~\ref{subsec:compare_ideal}, each referee generates labels from its corresponding kernel, an independent challenger predicts them, and the resulting performances are compared against the classical baselines described in Appendix~\ref{app:adversaries}.

\begin{figure}[t]
  \centering
  \includegraphics[width=\columnwidth]{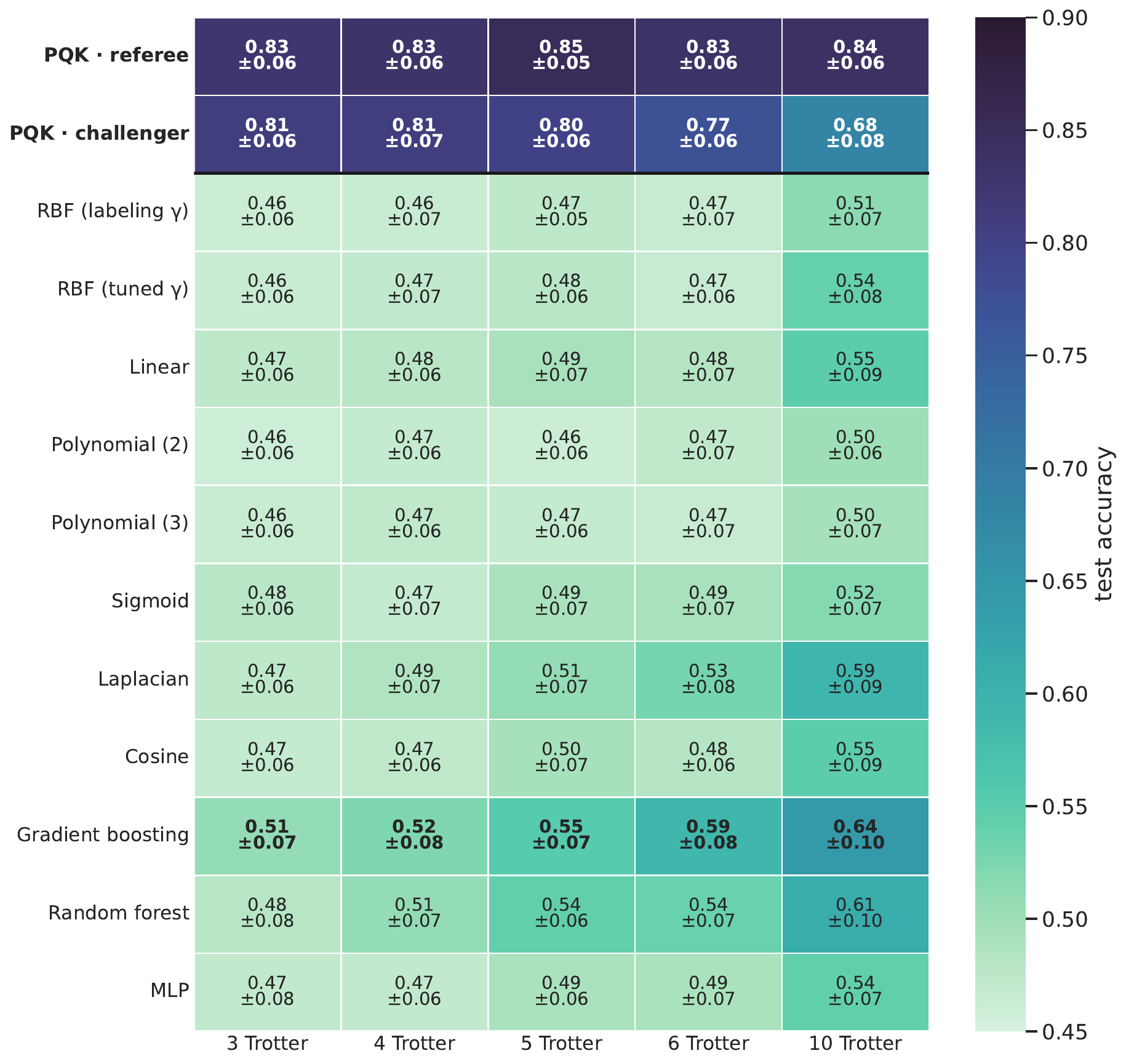}
  \caption{Mean test accuracy $\pm$ s.d.\ (Sec.~\ref{subsec:compare_ideal}) on engineered labels at $n{=}15$ for $T{=}3$--$10$. Rows: the referee on its own labels (\texttt{Emerald}), the challenger (independent run on \texttt{Garnet}), and the classical models of Appendix~\ref{app:adversaries}. As the depth (and therefore noise) increases, even though the referee's performance remains high, the quantum challenger's performance degrades whilst the classical challenger improves. This causes the gap in performance between quantum and classical to eventually disappear. The best classical challenger is shown in bold.}
  \label{fig:trotter_15q}
\end{figure}

At $n = 15$ qubits (Fig.~\ref{fig:trotter_15q}), self-evaluation by the referee yields consistent scores between $0.83$ and $0.85$ across all considered depths, confirming intrinsic label learnability. The challenger achieves comparable accuracy at shallow depths ($0.81$ at $T = 3$) but drops to $0.68$ at $T = 10$. The resulting gap to the leading classical model diminishes from $+0.30$ at $T = 3$ to $+0.04$ at $T = 10$, where the corrected resampled $t$-test no longer finds a gap, while it does for every depth up to $T = 6$ (Table~\ref{tab:gap_test}).

The comparison with the noiseless baseline shows what accompanies this degradation: hardware kernels deviate progressively from the ideal kernel with increasing $T$. Between $T = 3$ and $T = 10$, the CKA to the ideal kernel drops by about a third on both devices while $\varepsilon$ roughly triples (Table~\ref{tab:depth_ideal}, Appendix~\ref{app:unmit}). Because the ideal kernel preserves a separation of accuracy against the best classical baseline at $T = 10$ ($0.863$ vs.\ $0.588$ on labels engineered from the ideal kernel; Table~\ref{tab:em_main} caption), we conclude that the separation between quantum and classical predictors is attenuated on hardware primarily by accumulated device noise rather than as a result of the change in encoding caused by the number of Trotter steps.

Scaling the system to $n \in \{30, 45\}$ qubits (Appendix~\ref{app:unmit}, Fig.~\ref{fig:trotter_30q45q}), the two sizes behave very similarly: across all depths the gap stays within $[+0.17, +0.28]$, with a mean of $+0.18$ at $T = 10$ over both sizes, so it narrows far less with depth than at $n = 15$. The test finds a gap at every depth for both sizes (Table~\ref{tab:gap_test}); the lower bound for $95\%$ significance is smallest at $T = 10$ for $n = 45$ ($+0.065$), where the spread of the challenger's performance across runs is larger than at $n = 30$ ($0.09$ vs.\ $0.06$).

Unlike the cross-device benchmarking conducted at $15$ qubits, the challenger for $30$ and $45$ qubits constitutes an independent sub-chain embedded on the same quantum processor (Sec.~\ref{subsec:qpus}). How these pairs compare with the cross-device pair at $15$ qubits is discussed in Appendix~\ref{app:unmit}.

\begin{table}[t]
  \centering
  \caption{Gap between the quantum challenger and the best classical model, all tuned by cross-validation (CV). In parentheses, the 95\% lower confidence bound; fourth column: the gap against test-tuned classical models (Appendix~\ref{app:adversaries}); last column: the outcome of the corrected resampled $t$-test on the CV gap at the $5\%$ level. Each row reports the gap against the best classical adversary; in the last two blocks, each mitigated run is listed below the corresponding unmitigated baseline (the raw-counts row is identical to the $n{=}15$, $T{=}10$ row above). The gap shrinks with circuit depth and disappears at $n{=}15$, $T{=}10$, the regime in which we apply the error mitigation techniques. At $n{=}30$ and $45$, where referee and challenger are two chains on the same chip, it persists at every depth; at $n{=}15$, $T{=}10$, ZNE and APT restore the gap, while REM is not effective.}
  \label{tab:gap_test}
  \resizebox{\columnwidth}{!}{%
  \begin{tabular}{llccc}
    \toprule
    $n$ & $T$ / mitig. & gap (CV) & gap (test-tuned) & $t$-test \\
    \midrule
    15 & 3  & $+0.300$ ($\geq +0.238$) & $+0.250$ & gap \\
    15 & 4  & $+0.287$ ($\geq +0.247$) & $+0.237$ & gap \\
    15 & 5  & $+0.248$ ($\geq +0.190$) & $+0.205$ & gap \\
    15 & 6  & $+0.184$ ($\geq +0.101$) & $+0.159$ & gap \\
    15 & 10 & $+0.040$ ($\geq -0.070$) & $+0.017$ & no gap \\
    \midrule
    30 & 3  & $+0.281$ ($\geq +0.239$) & $+0.200$ & gap \\
    30 & 4  & $+0.246$ ($\geq +0.188$) & $+0.209$ & gap \\
    30 & 5  & $+0.224$ ($\geq +0.164$) & $+0.162$ & gap \\
    30 & 6  & $+0.228$ ($\geq +0.162$) & $+0.168$ & gap \\
    30 & 10 & $+0.181$ ($\geq +0.115$) & $+0.124$ & gap \\
    \midrule
    45 & 3  & $+0.245$ ($\geq +0.139$) & $+0.175$ & gap \\
    45 & 4  & $+0.242$ ($\geq +0.163$) & $+0.183$ & gap \\
    45 & 5  & $+0.256$ ($\geq +0.193$) & $+0.201$ & gap \\
    45 & 6  & $+0.242$ ($\geq +0.188$) & $+0.198$ & gap \\
    45 & 10 & $+0.170$ ($\geq +0.065$) & $+0.139$ & gap \\
    \midrule
    15 & $T{=}10$, none (original run) & $+0.059$ ($\geq -0.101$) & $+0.035$ & no gap \\
    15 & $T{=}10$, ZNE                 & $+0.139$ ($\geq +0.078$) & $+0.106$ & gap \\
    \midrule
    15 & $T{=}10$, none (raw counts)   & $+0.040$ ($\geq -0.070$) & $+0.017$ & no gap \\
    15 & $T{=}10$, REM                 & $+0.040$ ($\geq -0.091$) & $+0.015$ & no gap \\
    15 & $T{=}10$, APT                 & $+0.167$ ($\geq +0.105$) & $+0.137$ & gap \\
    \bottomrule
  \end{tabular}%
  }
\end{table}

\subsection{The effect of Error Mitigation}\label{subsubsec:em_results}

To determine whether error mitigation reduces the distance between the hardware and the ideal kernels, and how this affects the prediction gap, we analyse performance using $n = 15$ qubits, $T = 10$ Trotter steps, and the 600-datapoint pool, with the referee on \texttt{Emerald} and the challenger on \texttt{Garnet}. We do this as this is the first depth at which we do not observe a gap. Following the protocols described in Sec.~\ref{sec:methods}, we apply zero-noise extrapolation (ZNE), readout error mitigation (REM), and adaptive Pauli twirling (APT). The results of this section are summarised in Table~\ref{tab:em_main}, which compares the ideal and hardware kernels with different metrics; Table~\ref{tab:em_swap}, which separates the effect of the mitigation when creating the labels from that of the mitigation used by a quantum challenger; and the full prediction accuracies of all models are in Fig.~\ref{fig:em_heatmap} (Appendix~\ref{app:em}).

\begin{table*}[t]
  \centering
  \setlength{\tabcolsep}{4pt}
  \caption{Error mitigation on the QPU kernels at $n{=}15$, $T{=}10$, $N{=}600$, compared with the noiseless kernel on the same data points. $c$, $\lVert r\rVert$ and $\varepsilon$ characterise feature error and are defined in Sec.~\ref{subsec:compare_ideal}; $d_F$ is the relative Frobenius error to the ideal kernel. The last column is the mean accuracy (Sec.~\ref{subsec:compare_ideal}) on labels engineered from the ideal kernel. For comparison, the ideal kernel has a mean accuracy of $0.863$, and the best classical model $0.588$. Each mitigated row is compared with the unmitigated run at the top of its block. Entries $a\,/\,b$ are \texttt{Emerald} / \texttt{Garnet}. Bold: best value in each column for each device (highest $c$, CKA and accuracy; lowest $\lVert r\rVert$, $\varepsilon$ and $d_F$). APT is the only method to increase the accuracy, explained by the reduction in the effective error $\varepsilon$ and reflected in the larger centred kernel alignment (CKA).}
  \label{tab:em_main}
  \resizebox{\textwidth}{!}{%
  \begin{tabular}{llcccccc}
    \toprule
    Baseline & Mitig. & $c$ & $\lVert r\rVert$ & $\varepsilon$ & $d_F$ & CKA to ideal & acc.\ (ideal labels) \\
    \midrule
    \multirow{2}{*}{original run} & none & $0.552\,/\,0.392$ & $0.342\,/\,0.298$ & $0.619\,/\,0.760$ & $0.233\,/\,0.262$ & $0.563\,/\,0.488$ & $0.773\,/\,0.745$ \\
                                  & ZNE  & $\mathbf{0.825}\,/\,\mathbf{0.564}$ & $0.478\,/\,0.430$ & $0.580\,/\,0.762$ & $0.199\,/\,0.284$ & $0.633\,/\,0.503$ & $0.754\,/\,0.737$ \\
    \midrule
    \multirow{3}{*}{raw counts}   & none & $0.560\,/\,0.283$ & $0.326\,/\,0.314$ & $0.582\,/\,1.108$ & $0.194\,/\,0.360$ & $0.572\,/\,0.355$ & $0.768\,/\,0.701$ \\
                                  & REM  & $0.569\,/\,0.282$ & $0.331\,/\,0.338$ & $0.582\,/\,1.199$ & $0.196\,/\,0.334$ & $0.569\,/\,0.354$ & $0.770\,/\,0.703$ \\
                                  & APT & $0.684\,/\,0.400$ & $\mathbf{0.189}\,/\,\mathbf{0.193}$ & $\mathbf{0.276}\,/\,\mathbf{0.483}$ & $\mathbf{0.139}\,/\,\mathbf{0.191}$ & $\mathbf{0.903}\,/\,\mathbf{0.700}$ & $\mathbf{0.838}\,/\,\mathbf{0.763}$ \\
    \bottomrule
  \end{tabular}%
  }
\end{table*}

\begin{table}[t]
    \centering
    \caption{Mean accuracy (Sec.~\ref{subsec:compare_ideal}) of the challenger (\texttt{Garnet}) on labels engineered by each referee (\texttt{Emerald}); baselines as in Table~\ref{tab:em_main}. The off-diagonal entries separate the effect of using mitigation to produce the labels from that of mitigating the challenger. Bold: highest accuracy for each method. For both methods, either party using mitigation improves the accuracy of the challenger.}
    \label{tab:em_swap}
    \begin{tabular}{llcc}
      \toprule
      Mitig. & challenger & unmitigated & mitigated \\
       &  & referee & referee \\
      \midrule
      \multirow{2}{*}{ZNE}        & unmitigated & $0.700$ & $0.740$ \\
                                  & mitigated   & $0.715$ & $\mathbf{0.748}$ \\
      \midrule
      \multirow{2}{*}{APT} & unmitigated & $0.683$ & $0.687$ \\
                                  & mitigated   & $0.717$ & $\mathbf{0.761}$ \\
      \bottomrule
    \end{tabular}
\end{table}

\textbf{The effect of mitigation on features}: Error mitigation is expected to shift measured observables toward their ideal values, as observed with ZNE: it increases the scaling factor $c$ by $44\%\text{--}49\%$ and reduces the relative feature error by approximately $10\%$ for both devices ($0.563 \to 0.509$ on \texttt{Emerald}, $0.677 \to 0.612$ on \texttt{Garnet}). However, extrapolation simultaneously amplifies the residual by a similar proportion (an increase in $\lVert r\rVert$ of $40\%\text{--}44\%$; see Table~\ref{tab:em_main}). By construction, REM can only adjust $c$ within the readout ceiling (Sec.~\ref{subsec:em_methods}), yielding $+1.6\%$ out of a potential $+2.1\%$ on \texttt{Emerald}, and $-0.5\%$ against a potential $+3.6\%$ on \texttt{Garnet}. APT instead increases $c$ while also suppressing the residual norm: the relative feature error drops by more than ZNE achieves ($0.547 \to 0.368$ on \texttt{Emerald}, $0.782 \to 0.630$ on \texttt{Garnet}), and the effective error $\varepsilon$ roughly halves on both devices.

\textbf{The effect of mitigation on kernels}: Moving closer to the ideal features does not necessarily mean moving closer to the ideal kernel. The bandwidth in Eq.~\eqref{eq:bandwidth} is normalized by the variance of the features. Consequently, if every feature shrinks by the same factor, $f_{\mathrm{hw}} = c\,f$, their standard deviation shrinks by the same factor. As the effective bandwidth adjusts to compensate, the kernel remains unchanged:

\begin{multline}
  \gamma_{\mathrm{eff}}^{\mathrm{hw}}\,\lVert f_{\mathrm{hw}}(\mathbf{x})-f_{\mathrm{hw}}(\mathbf{x}')\rVert^2 = \gamma_{\mathrm{eff}}\,\lVert
f(\mathbf{x})-f(\mathbf{x}')\rVert^2 \\
  \;\Longrightarrow\; K_{\mathrm{hw}} = K .
  \label{eq:invariance}
\end{multline}

The kernel is therefore invariant under uniform contractions. Decomposing the measured features in Eq.~\eqref{eq:split} as $c\,[f(\mathbf{x}) + r(\mathbf{x})/c]$, the leading factor $c$ cancels out; the hardware kernel thus depends on the ideal features plus an effective residual $r/c$, whose magnitude is $\varepsilon$ (Eq.~\eqref{eq:effnoise}). Raising $c$ towards $1$, which is the primary goal of ZNE, alters the kernel solely through its net effect on the ratio $r/c$.

\begin{figure}[t]
    \centering
    \includegraphics[width=\columnwidth]{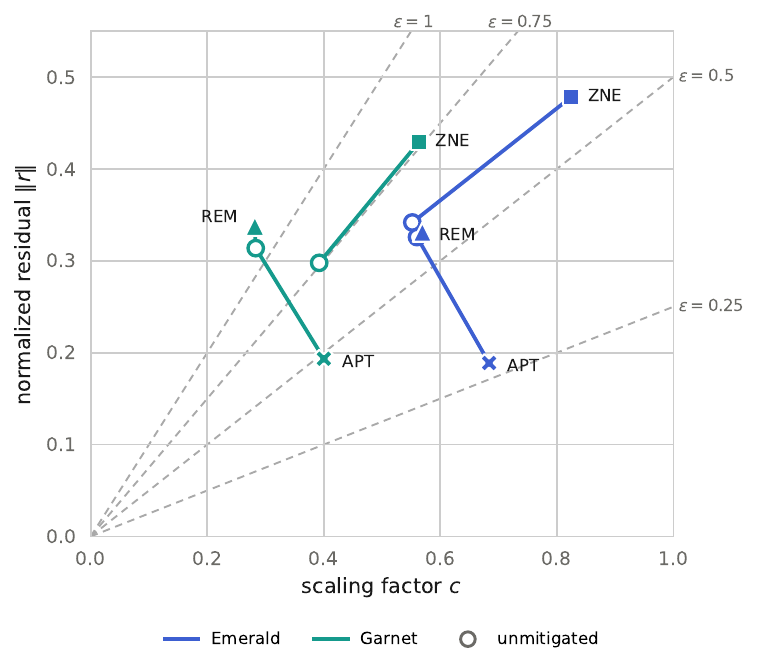}
    \caption{Hardware kernels on the $(c, \lVert r\rVert)$ plane, from Table~\ref{tab:em_main}. Dashed lines: constant $\varepsilon = \lVert r\rVert/c$, the size of the effective residual that sets the kernel's distance to the ideal (Eq.~\eqref{eq:invariance}). Each line joins an unmitigated baseline (hollow circle) to its mitigated kernel (baselines as in Table~\ref{tab:em_main}). ZNE moves both kernels almost along the dashed lines, leaving $\varepsilon$ nearly unchanged; APT moves them across, roughly halving $\varepsilon$.}
    \label{fig:em_plane}
\end{figure}

This is precisely what we observe. Because ZNE scales $c$ and $\lVert r\rVert$ by similar amounts, it moves the kernels almost along a line of constant $\varepsilon$ (Fig.~\ref{fig:em_plane}), and the kernels shift only slightly toward the ideal (Table~\ref{tab:em_main}). An unchanged $\varepsilon$ does not, however, mean an unchanged kernel. $\varepsilon$ measures how much noise is left relative to the signal, not which noise: ZNE extrapolates each feature separately, and replaces the leftover noise with a different one of about the same size. The kernels therefore change considerably (the CKA between each unmitigated kernel and its ZNE version is only $0.552$ for \texttt{Emerald} and $0.632$ for \texttt{Garnet}), yet end up only slightly closer to the ideal, because the new noise is barely smaller than the original; the kernel changes in a direction largely tangential to that pointing towards the ideal (Appendix~\ref{app:em_geometry}). What $\varepsilon$ predicts is the distance to the ideal, not how much the kernel changes overall: the ZNE \texttt{Garnet} kernel has a higher $c$ than the unmitigated \texttt{Emerald} kernel but also a larger $\varepsilon$, and it lies further from the ideal (Appendix~\ref{app:em_rank}). By relative Frobenius error, ZNE even moves the \texttt{Garnet} kernel away from the ideal, because $d_F$ also counts constant offsets and uniform rescaling of the kernel entries, which the SVM and the CKA ignore. APT, by contrast, moves the kernels across the lines of constant $\varepsilon$ (Fig.~\ref{fig:em_plane}) and, as predicted by Eq.~\eqref{eq:invariance}, much closer to the ideal: the CKA increases from $0.572$ to $0.903$ on \texttt{Emerald} and from $0.355$ to $0.700$ on \texttt{Garnet}, and, unlike for ZNE, the relative Frobenius error to the ideal decreases on both devices (Table~\ref{tab:em_main}). Finally, REM preserves the kernel structure entirely: its CKA relative to the unmitigated kernel from the raw counts is $1.000$ for both devices.

\textbf{The effect of mitigation on prediction}: If mitigation brings the kernel closer to the ideal, one could expect a larger gap. We first apply each method to both parties, as in a full experiment: the referee engineers the labels from its mitigated kernel, and the challenger predicts them with its mitigated kernel; the baseline is the same pair without mitigation. With ZNE on both, the challenger's gap to the best classical model increases and, unlike in the unmitigated run, passes the test; with REM on both it stays without a gap; and with APT on both it widens, passing the test with the largest lower bound among the $15$-qubit runs at $T = 10$ (Table~\ref{tab:gap_test}). The referee's gap on its own labels also grows with ZNE, but solely because classical model performance degrades on the ZNE labels (Fig.~\ref{fig:em_heatmap}); the referee's own accuracy remains virtually unchanged.

Table~\ref{tab:em_swap} separates the effect of mitigation on each party individually.
For ZNE, the labels of the ZNE referee account for most of the challenger's $+0.048$ gain ($+0.033$ to $+0.040$, depending on the order one sums the contribution of mitigation on each party),
whereas running ZNE on the challenger contributes only $+0.008$ to $+0.015$: the increase in the gap under ZNE thus stems mainly from the new labels after mitigation rather than from the mitigation by the challenger. We call the improvement in accuracy of an unmitigated challenger on labels produced after mitigation over the unmitigated labels the \emph{label effect}. For our ZNE experiments, it is not spread evenly over the ten subsamples: most of it comes from two, in which the ZNE referee selects a different labelling bandwidth $\gamma_Q^*$ (Sec.~\ref{subsec:hyperparams}) than the unmitigated referee. Regenerating the ZNE labels in those two subsamples with the unmitigated referee's bandwidth roughly halves the label effect (Appendix~\ref{app:em_labels}), therefore about half of it comes from this change of bandwidth, and the rest from the effect of ZNE on the kernel itself. For APT, the label effect alone adds almost nothing ($+0.004$ with an unmitigated challenger): the challenger's $+0.078$ gain appears only when the challenger itself is mitigated, which accounts for $+0.034$ to $+0.074$ of it.

Labels engineered from the ideal noiseless kernel cannot exhibit the label effect, and they tell a different story (Table~\ref{tab:em_main}, last column). Applying ZNE to \texttt{Emerald} leads to slightly worse predictions than the unmitigated kernel ($0.773 \to 0.754$) and brings no improvement for \texttt{Garnet}; REM leaves the accuracy virtually unchanged, while APT improves it on both devices, bringing \texttt{Emerald}'s accuracy within $0.025$ of the ideal kernel's; the ideal kernel itself ($0.863$) stays ahead of every hardware kernel. When predicting its own labels, the ideal kernel also maintains the largest gap ($+0.275$), larger than any hardware gap. Being closer to the ideal is not enough to improve accuracy either: the ZNE \texttt{Emerald} kernel has a higher CKA to the ideal than the unmitigated one, yet predicts the ideal labels worse. ZNE therefore increases the gap by changing the task the referee engineers, not by improving the kernel: mitigating the challenger adds little improvement on the hardware labels (Table~\ref{tab:em_swap}). When predicting the ideal labels, ZNE does not improve the accuracy of either hardware kernel (Table~\ref{tab:em_main}). 

Without an ideal reference, the challenger's higher accuracy and larger gap on the labels engineered by the ZNE referee would suggest that ZNE improves the quantum kernel. ZNE and APT thus both open a gap for hardware-generated engineered labels, but only APT does so by making the kernel itself better.

\subsection{Classical challenger performance using a matrix product state kernel on hardware-produced data}\label{subsec:mps}

As discussed in Sec.~\ref{sec:intro}, a full advantage experiment requires the kernel to be classically hard to compute; in our case, the data-encoding circuits are still classically simulable (see Sec.~\ref{subsec:encoding}).
Here we present the results for a challenger kernel calculated using a Matrix Product State (MPS) approach, one of the standard methods used to simulate quantum states \cite{VidalMPS,PGMPS}. Although MPS can represent any quantum state, efficient approximations work particularly well for one dimensional systems and when the entanglement generated by the quantum circuit is not too high, making them a natural candidate for approximating our states. 

To demonstrate this, we take the most challenging experiment reported, which is the one with the highest number of qubits ($n = 45$) and the largest depth ($T=10$), and use an MPS simulator with max bond dimension $\chi=128$ to compute a challenger kernel. This is then used to predict the labels engineered with a quantum kernel computed on the QPU (without error mitigation). The classically simulated kernel achieves comparable prediction accuracy as a quantum challenger evaluated on hardware ($0.72 \pm 0.10$ - see Fig.~\ref{fig:trotter_45q} for comparison with the other challengers), and thus gives a comparable prediction gap too with respect to the best circuit-agnostic classical adversary ($+0.161$ with cross-validation applied). 

These results confirm that even the biggest experiment here reported is still within reach of classical simulations, and that quantum advantage will require quantum kernels designed to inhibit state-of-the-art simulation methods.
Since our matrix product approach was able to simulate each circuit in under 20 seconds and within an error of $10^{-2}$, we did not further optimise our simulation approach, such as restricting the tensor network to the corresponding light cone of each local observable \cite{VidalLC,MarkovShi}. It is surprising to see that these high accuracy simulations do not outperform a quantum challenger, with the corrected $t$-test giving no significant difference between the two. This raises the question of whether noise should be added to the simulation, to try and better mimic the referee in order to improve accuracy.

\subsection{Producing new test points}
\label{subsec:newlabels}

One limitation of the engineered dataset construction presented in \cite{huang2021power} and described in Sec.~\ref{subsec:learning_problem} and \ref{subsec:hyperparams} is the use of the test data in the construction. In fact, for a fixed $N = N_{\mathrm{train}}+N_{\mathrm{test}}$, the engineered dataset generation creates a dataset of size $N$, which is then split again into train and test. 
This may lead to an inflated predictive performance on the test set as the training labels are generated using also the test data, and, most importantly, it is not possible to calculate the performance of the quantum kernel on a new data point not included in $N$, as this will not have an assigned label. We investigate here the robustness of the predictive advantage of the quantum kernel on an engineered learning problem constructed via a function that uses only the training data to engineer the training labels and can generalise to new test points.

Recall that the labels in \cite{huang2021power} are generated via the transformation:
\begin{equation}
y_{i} = \sum_{j} \sqrt{K_Q}_{ij}v_j,
\end{equation}
where $v$ is the eigenvector corresponding to the eigenvalue $g^2$ in Eq.~\eqref{ggen}. Assuming that for a new data point $\mathbf{x}_k$ the kernel elements $(K_{Q})_{ki}:= k(\mathbf{x}_k,\mathbf{x}_i)$ have been calculated, we define the function:
\begin{equation}
y_{k} := \sum_{i,j} (K_{Q})_{ki}(\sqrt{K_Q})^{-1}_{ij}v_j,
\label{eq:newlabels}
\end{equation}
which allows us to generate the label for a new test point. To see if with this new function we can reproduce a performance gap between a quantum and a classical challenger, we repeat the $15$ qubit experiment presented in Sec.~\ref{subsubsec:unmit_results}, using the same sub-samples of data. The results for the test accuracy are shown in Fig.~\ref{fig:newlabels} (Appendix~\ref{app:unmit}), where we only report the best found classical model. We observe that the performance is lower both for the referee and the challenger quantum kernels, confirming that the inclusion of the test dataset in the label construction inflates the predictive performance of the PQK. 
Nevertheless, a gap in performance is present also for the new engineered labels, as reported in Table \ref{tab:gap_test_newlabels}. We find that the predictive performance gap is smaller than with the previous construction, but still persists under QPU noise. The corrected $t$-test confirms that the advantage over the classical model is lost at $T=10$, when no error mitigation is applied.
\begin{table}
\caption{Gap between the quantum challenger and the best classical model, all tuned by cross-validation (CV), on the new engineered labels via Eq.~\eqref{eq:newlabels}. In parentheses, the 95\% lower confidence bound; fourth column: the gap against test-tuned classical models (Appendix~\ref{app:adversaries}); last column: the outcome of the corrected resampled $t$-test on the CV gap at the $5\%$ level. The values of the gap are smaller than the ones obtained with the construction of \cite{huang2021power}. Nevertheless, the test confirms analogous behaviour, with the gap disappearing at $T=10$.}
\label{tab:gap_test_newlabels}
  \begin{tabular}{llccc}
    \toprule
    $n$ & $T$ & gap (CV) & gap (test-tuned) & test \\
    \midrule
    15 & 3  & $+0.215$ ($\geq +0.144$) & $+0.152$ & gap \\
    15 & 4  & $+0.193$ ($\geq +0.159$) & $+0.125$ & gap \\
    15 & 5  & $+0.122$ ($\geq +0.080$) & $+0.052$ & gap \\
    15 & 6  & $+0.108$ ($\geq +0.040$) & $+0.036$ & gap \\
    15 & 10 & $+0.007$ ($\geq -0.061$) & $-0.041$ & no gap \\
    \bottomrule
  \end{tabular}%
\end{table}

These results show that it is possible to generalise the engineered learning problem to label any new test point, with the PQK still giving an advantage. The limitations coming from the hardware noise impact the two different constructions equivalently; however, as including test points in the labelling process does result in larger gaps, this may still be the better candidate to use in a boundary pushing advantage experiment.

\subsection{The effect of data size}
\label{subsec:number_of_data}

All the results presented in \cite{huang2021power} hold in an ideal scenario, when the quantum kernel is not affected by noise. As a consequence, the prediction error scales as $1/\sqrt{N_{\mathrm{train}}}$, with $N_{\mathrm{train}}$ the number of training data. In \cite{wang2021towards} it was shown that when depolarizing and sampling noise are considered, the error can actually increase with the number of data. In other words, for a fixed number of shots there is an optimal number of data $N_{\mathrm{train}}$ to use to achieve maximal performance, after which the performance starts decreasing. More precisely, the generalization error bound has, with respect to the ideal one, an additional term that scales as $N_{\mathrm{train}}^{1/4}$. The result in \cite{wang2021towards} is derived using fidelity kernels and a simplified version of noise model with respect to current QPUs, thus it is not directly applicable to the case studied in this work. Nevertheless, we can empirically observe how the prediction accuracy changes with the number of data. We will do this with two different experiments, which answer the following two questions:

\textbf{For the referee experiment}. \textit{Is there an optimal $N = N_{\mathrm{train}}+N_{\mathrm{test}}$ to use to setup the learning problem?} Here optimal refers to the biggest possible performance gap with respect to a classical adversary. Note that in this case the protocol used to construct the engineered learning problem is the one described in Sec.~\ref{subsec:hyperparams}.

\textbf{For the challenger experiment}. \textit{Is there an optimal $N_{\mathrm{train}}$ to use to predict a given set of labels?} In this case, the test set and labels are fixed.

\begin{figure}[tb]
    \centering
    \includegraphics[width=\columnwidth]{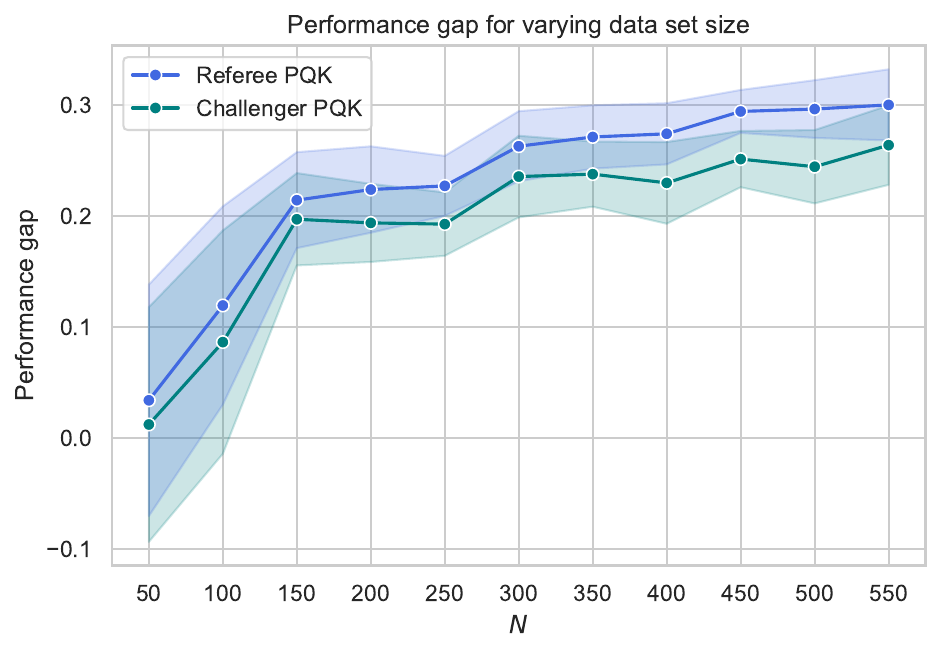}
    \caption{Prediction accuracy improvement against the best classical baseline as a function of the number $N$ of data used to construct the engineered labels. The classical adversaries are test-tuned (see Appendix~\ref{app:adversaries}). For both projected quantum kernels (PQKs), the performance gap keeps increasing with increasing number of data.}
    \label{fig:referee_exp}
\end{figure}

\begin{figure}[tb]
    \centering
    \includegraphics[width=\columnwidth]{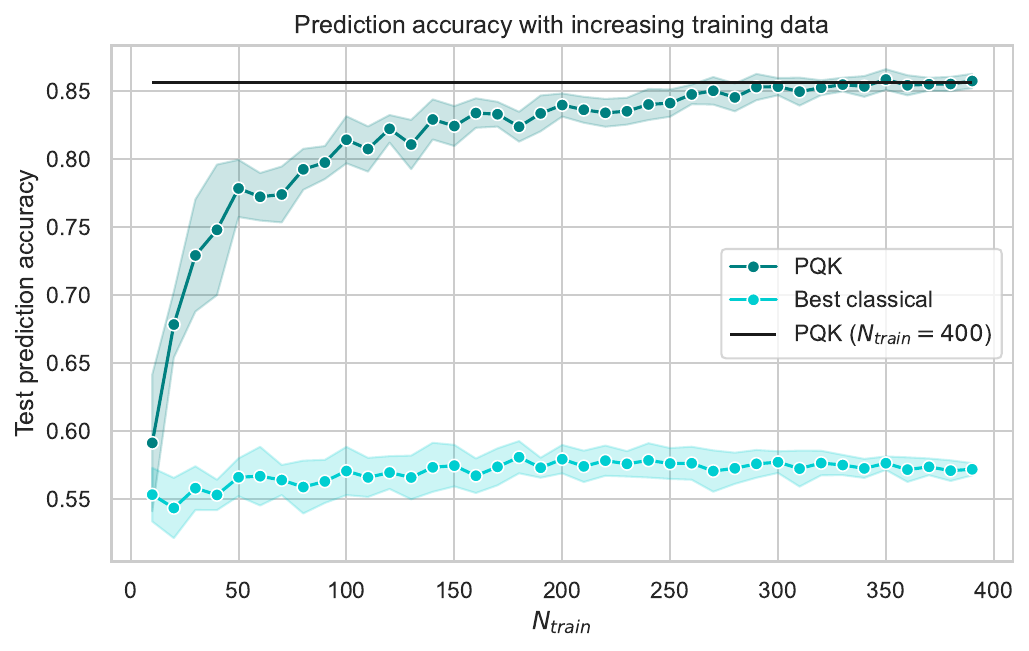}
    \caption{Test prediction accuracy of the challenger quantum kernel and the best classical adversary for increasing $N_{\mathrm{train}}$. The classical adversaries are test-tuned (see Appendix~\ref{app:adversaries}). The PQK accuracy plateaus before reaching the maximum amount of training data available to the challenger.}
    \label{fig:chal_exp}
\end{figure}

For these exploratory experiments, we limit the number of qubits to $15$ and set $N_{\mathrm{max}}=600$. As with the other experiments, we use quantum kernels evaluated on different QPUs and resampling to get average prediction performance over $10$ sub-samples of different sizes.
We present the results for the referee experiment in Fig.~\ref{fig:referee_exp}, where the performance gap of the two quantum kernels against the best classical adversary is shown for increasing $N$. The train-test split is fixed to $2/3$ train and $1/3$ test. We see that for both quantum kernels the gap increases when adding more data: we do not observe an optimal peak in the interval considered. 
If instead we use all data to engineer the labels, but vary the number $N_{\mathrm{train}}$ available to the challengers, we obtain the results shown in Fig.~\ref{fig:chal_exp}. Here we report the performance of the challenger quantum kernel and the best classical adversary. All models predict on the same $N_\mathrm{test}=200$ test labels; we find that, up to $N_{\mathrm{train}}=250$, adding more data in the training helps improving the test prediction accuracy of the PQK. For $N_{\mathrm{train}}>250$, the improvement in accuracy becomes much smaller and plateaus around the model performance achieved with $N_{\mathrm{train}}=400$. The classical prediction accuracy is far less sensitive to the number of data used in the train set, and stays well below the quantum accuracy for all values of $N_{\mathrm{train}}$.
We do not find in these results the behaviour expected from the bound described in \cite{wang2021towards}. This could be due to several reasons: the optimal number of data is bigger than $N_{\mathrm{max}}$, the scaling of the error bound is different under the QPU noise or when using PQKs instead of fidelity kernels. Further investigations are thus needed to assess the correct scaling of the error bound for the experiments presented in this work.

\section{Discussion and Conclusion}
\label{sec:discussion}
In this work, we address the question: \textit{does a performance gap in prediction exist between a quantum kernel and a classical adversary when the quantum kernel is computed on current hardware?} We show that for different qubit numbers and circuit depths, we are able to demonstrate a performance gap between a quantum and the best classical challenger chosen from a suite of standard classical machine learning techniques. To obtain these results, we conduct a thorough optimization of the bandwidth $\gamma$ for all the kernels involved in the construction of the learning problem and the ones used for prediction. Importantly, we show reproducibility of the results on different hardware; our quantum kernels for the referee engineering the problem and the challenger attempting to solve it are calculated either on separate QPUs, or using layouts on the same QPU designed to have different noise profiles. In contrast to studying ideal quantum kernels, this reproducibility study is needed due to the risk of fitting to the noise of a specific device when constructing the learning problem.

While scaling the number of qubits does not change the results significantly, the effect of hardware noise can be seen when increasing the depth of the data-encoding quantum circuits (equivalent to increasing the number of Trotter steps T for our case). At $T=10$ and $n=15$, the gap disappears and the quantum challenger no longer outperforms the classical one. We have thus investigated whether error mitigation can help revive the predictive accuracy gap. 

Among the techniques evaluated in this work, we find REM to be the least effective: its impact is bounded by a readout error of a few percent and leaves the kernel practically untouched, failing to alter the performance gap between quantum and classical methods on labels engineered using hardware, or to improve the accuracy on labels created with an ideal noiseless kernel. ZNE as a mitigation strategy is able to successfully restore the quantum-classical performance gap, but does so by modifying the task to be more difficult for the classical methods; mitigating the quantum challenger itself adds little on hardware-engineered labels, and neither hardware kernel improves on labels created with the noiseless kernel. This can be explained because ZNE amplifies the residual feature error roughly as much as it restores the signal, nearly shifting the kernel along a contour of constant effective error $\varepsilon$ (Fig.~\ref{fig:em_plane}). 
By contrast, APT suppresses the residual directly and roughly halves $\varepsilon$; this moves the resulting kernels significantly closer to the ideal, resulting in more accurate prediction of the ideal labels, and restoring the quantum-classical performance gap by improving the performance of the quantum challenger.
Consequently, while error mitigation can indeed revive the performance gap, comparison with the ideal kernel allows us to understand where this improvement comes from: with ZNE the impact comes from the effect on the task defined by the referee, whereas with APT the benefit comes from a genuine improvement of the quantum challenger. This information can help to inform future experimental design; such as allocating resources, or if some mitigation methods are not available to all devices.

Another contribution of this study is the generalisation of the labels construction protocol to a function which can be used to generate a label for any new test point and does not restrict the problem to $N$ fixed data points. This approach offers more flexibility, but reduces the size of the performance gap; if the aim is solely to push the discrepancy between quantum and classical performance to the largest value, then the approach limited to a fixed number of data $N$ is likely preferable. We also explored the role of the size of $N$ on the prediction accuracy; as the scaling of the prediction error with $N$ was central in the results of \cite{huang2021power}. Here, motivated by the results of \cite{wang2021towards}, we wanted to empirically determine if an optimal $N$ exists under QPU noise. Our results show that, in the range of dataset sizes considered here, the performance gap keeps increasing when adding more data to the problem construction, suggesting that a referee should use the maximum number of data available. On the other hand, when a quantum challenger wants to solve the task, the number of training data they use could be reduced, as the prediction accuracy plateaus before the maximum $N_{\mathrm{train}}$ is reached. This implies resource-constrained quantum challengers may be better off investing in error mitigation for a subset of the data available to them. Further investigations on the role of noise remain to be done, and the derivation of a prediction error bound tailored to our specific setup remains open.

Finally, we have presented a classically simulated quantum kernel to ultimately clarify that the experiments conducted in this study are not quantum advantage experiments. We used MPS simulations to calculate the kernel with $45$ qubits and $10$ Trotter steps; the predictive performance of this kernel on the problem engineered by a quantum kernel evaluated on the QPU is comparable to the challenger also using quantum hardware. The observation that the classically simulated kernel does not outperform the (noisy) hardware raises the interesting question of whether adding noise to the simulations would actually be beneficial in this case to achieve higher accuracies.
In contrast to fidelity kernels, where the overlap is computed directly on a quantum computer, projected kernels can be more fairly characterised as the extraction of a set of quantum features, followed by the application of a purely classical kernel function. It would be interesting to understand whether the addition of the kernel function can be dropped from the learning problem construction or whether it brings tangible benefits, such as the invariance under shrinking noise as described in Sec.~\ref{subsubsec:em_results}. It would also be worthwhile to investigate further if the label  engineering approach in \cite{SakhnenkoArtificial} produces larger gaps than the approach used here.

From the results here we conclude that, if constructing labels using a kernel evaluated on quantum hardware, the best classical challengers will be those that try to simulate the circuits used directly, rather than the standard classical models. Therefore, the logical next step is to replace the embedding chosen in this work by one informed by the current discourse about where classical simulation of quantum systems starts to break down. Although earlier proposed applications for quantum kernels \cite{Many_Body_Problems,Phases_Support,Phases_3} relied on the difficulty of ground state preparation to produce a quantum advantage, the authors understand that time evolution of quantum systems is seen as a more promising area for quantum computers to overtake classical methods, as supported by recent works \cite{IQM_Utility,Andersen2025,Shaw2024, Abanin2025, bandyopadhyay2026provable}.

\section*{Data and Code Availability}
The projected features and kernel matrices from all hardware and simulator runs, together with the engineered labels, train/test splits and test accuracies underlying the main results of this work, are available on Zenodo~\cite{zenodo_data}, and the code to reproduce the main results from these data on GitHub~\cite{github_repo}.

\section*{Acknowledgements}
The authors thank Alessio Calzona and Vicente Pina for their help in understanding and implementing their adaptive Pauli twirling error suppression method, the details of which are soon to be released in an upcoming publication. The authors also thank Aniket Rath, for his comments on the proper usage of the IQM Qubit Selector \cite{iqm_qubit_selector}.

E.P. and G.B.S. thank the Région Île-de-France under the Pack Quantique programme (Grant No.~23002667), whose support enabled their internships at IQM France SAS, during which they obtained their knowledge of quantum kernels, and the discrepancy between mitigation for features and kernels was first observed.
\subsection*{AI Disclosure Statement}
Claude Opus was used as an assistant to polish and refine the language used in this paper, including grammar, wording and consistency across this work. Additionally it also assisted in the creation of figures, and validation of experimentation setup. Every piece of AI-generated content was reviewed, verified, and edited by the authors, who take full responsibility for the final manuscript.

\bibliographystyle{apsrev4-2}
\bibliography{reference}

@article{huang2021power,
  title={Power of data in quantum machine learning},
  author={Huang, Hsin-Yuan and Broughton, Michael and Mohseni, Masoud and Babbush, Ryan and Boixo, Sergio and Neven, Hartmut and McClean, Jarrod R},
  journal={Nature Communications},
  volume={12},
  number={1},
  pages={2631},
  year={2021},
  publisher={Nature Publishing Group}
}

@article{advantageSweke,
    author = {Sweke, Ryan and Seifert, Jean-Pierre  and Hangleiter,Dominik and Eisert, Jens},
    title = {On the Quantum versus Classical Learnability of Discrete Distributions},
    journal = {Quantum},
    pages = {417},
    issue = {5},
    year = {2021}
}

@article{Phases_3,
  title={Quantum Phase Recognition via Quantum Kernel Methods},
  author={Yusen Wu and Bujiao Wu and Jingbo Wang and Xiao Yuan},
  journal={Quantum},
  volume={7},
  pages={981},
  year={2023},
}

@article{QASchuld,
  title = {Is Quantum Advantage the Right Goal for Quantum Machine Learning?},
  author = {Schuld, Maria and Killoran, Nathan},
  journal = {PRX Quantum},
  volume = {3},
  issue = {3},
  pages = {030101},
  numpages = {13},
  year = {2022},
  month = {Jul},
  publisher = {American Physical Society},
  doi = {10.1103/PRXQuantum.3.030101},
  url = {https://link.aps.org/doi/10.1103/PRXQuantum.3.030101}
}

@misc{BenchmarkBowles,
      title={Better than classical? The subtle art of benchmarking quantum machine learning models}, 
      author={Joseph Bowles and Shahnawaz Ahmed and Maria Schuld},
      year={2024},
      eprint={2403.07059},
      archivePrefix={arXiv},
      primaryClass={quant-ph},
      url={https://arxiv.org/abs/2403.07059}, 
}

@inproceedings{NeuralTangent,
author = {Jacot, Arthur and Gabriel, Franck and Hongler, Cl{\'e}ment},
title = {Neural tangent kernel: convergence and generalization in neural networks},
year = {2018},
publisher = {Curran Associates Inc.},
address = {Red Hook, NY, USA},
booktitle = {Proceedings of the 32nd International Conference on Neural Information Processing Systems},
pages = {8580–8589},
numpages = {10},
location = {Montr{\'e}al, Canada},
series = {NIPS'18}
}

@article{kubler2021inductive,
  title={The inductive bias of quantum kernels},
  author={K{\"u}bler, Jonas and Buchholz, Simon and Sch{\"o}lkopf, Bernhard},
  journal={Advances in Neural Information Processing Systems},
  volume={34},
  pages={12661--12673},
  year={2021}
}

@article{Yin_Photon,
  title={Experimental quantum-enhanced kernel-based machine learning on a photonic processor},
  author={Zhenghao Yin and Iris Agresti and Giovanni de Felice and Douglas Brown and Alexis Toumi and Ciro Pentangelo and Simone Piacentini and Andrea Crespi and Francesco Ceccarelli and Roberto Osellame and Bob Coecke and Philip Walther},
  journal={Nature Photonics},
  volume={19},
  pages={1020–1027},
  year={2025}
}

@article{Arute_2019,
    author={Arute, Frank and Arya, Kunal and Babbush, Ryan and Bacon, Dave and Bardin, Joseph C. and Barends, Rami and Biswas, Rupak and Boixo, Sergio and Brandao, Fernando G. S. L. and Buell, David A. and Burkett, Brian and Chen, Yu and Chen, Zijun and Chiaro, Ben and Collins, Roberto and Courtney, William and Dunsworth, Andrew and Farhi, Edward and Foxen, Brooks and Fowler, Austin and Gidney, Craig and Giustina, Marissa and Graff, Rob and Guerin, Keith and Habegger, Steve and Harrigan, Matthew P. and Hartmann, Michael J. and Ho, Alan and Hoffmann, Markus and Huang, Trent and Humble, Travis S. and Isakov, Sergei V. and Jeffrey, Evan and Jiang, Zhang and Kafri, Dvir and Kechedzhi, Kostyantyn and Kelly, Julian and Klimov, Paul V. and Knysh, Sergey and Korotkov, Alexander and Kostritsa, Fedor and Landhuis, David and Lindmark, Mike and Lucero, Erik and Lyakh, Dmitry and Mandr, Salvatore and McClean, Jarrod R. and McEwen, Matthew and Megrant, Anthony and Mi, Xiao and Michielsen, Kristel and Mohseni, Masoud and Mutus, Josh and Naaman, Ofer and Neeley, Matthew and Neill, Charles and Niu, Murphy Yuezhen and Ostby, Eric and Petukhov, Andre and Platt, John C. and Quintana, Chris and Rieffel, Eleanor G. and Roushan, Pedram and Rubin, Nicholas C. and Sank, Daniel and Satzinger, Kevin J. and Smelyanskiy, Vadim and Sung, Kevin J. and Trevithick, Matthew D. and Vainsencher, Amit and Villalonga, Benjamin and White, Theodore and Yao, Z. Jamie and Yeh, Ping and Zalcman, Adam and Neven, Hartmut and Martinis, John M.},
    title = {Quantum supremacy using a programmable superconducting processor},
    publisher = {Nature Publishing Group},
    journal = {Nature},
    pages = {505--510},
    month = {oct},
    volume = {574},
    year = {2019},
    doi = {10.1038/s41586-019-1666-5},
}

@article{Freq_Kernels,
    
AUTHOR={Nader, Andrew  and Dubois, Marc-André  and Kundur, Deepa },
           
TITLE={Exploring quantum learning in the smart grid through the evolution of noisy finite fourier series},
          
JOURNAL={Frontiers in Energy Research},
          
VOLUME={11},
  
YEAR={2023},
  
URL={https://www.frontiersin.org/journals/energy-research/articles/10.3389/fenrg.2023.1061602},
  
DOI={10.3389/fenrg.2023.1061602},
  
ISSN={2296-598X},
  
}

@article{Gupta_2025,
title={Potential of quantum machine learning for processing multispectral Earth observation data},
volume={73}, 
url={https://wydawnictwo.pan.pl/index.php/bpasts/article/view/854}, 
DOI={10.24425/bpasts.2025.154279}, 
number={5}, 
journal={Bulletin of the Polish Academy of Sciences Technical Sciences}, 
author={Gupta, Manish K. and Romaszewski, Michał and Gawron, Piotr}, 
year={2025}, 
month={Aug.}, 
}

@article{BPEquivalence,
  title = {Equivalence between exponential concentration in quantum machine learning kernels and barren plateaus in variational algorithms},
  author = {Kairon, Pranav and J\"ager, Jonas and Krems, Roman V.},
  journal = {Phys. Rev. A},
  volume = {114},
  issue = {2},
  pages = {022417},
  numpages = {10},
  year = {2026},
  month = {Aug},
  publisher = {American Physical Society},
  doi = {10.1103/6hbr-kj64},
  url = {https://link.aps.org/doi/10.1103/6hbr-kj64}
}

@article{Graph_Data,
  title = {Quantum feature maps for graph machine learning on a neutral atom quantum processor},
  author = {Albrecht, Boris and Dalyac, Constantin and Leclerc, Lucas and Ortiz-Guti\'errez, Luis and Thabet, Slimane and D'Arcangelo, Mauro and Cline, Julia R. K. and Elfving, Vincent E. and Lassabli\`ere, Lucas and Silv\'erio, Henrique and Ximenez, Bruno and Henry, Louis-Paul and Signoles, Adrien and Henriet, Lo\"{\i}c},
  journal = {Phys. Rev. A},
  volume = {107},
  issue = {4},
  pages = {042615},
  numpages = {16},
  year = {2023},
  month = {Apr},
  publisher = {American Physical Society},
  doi = {10.1103/PhysRevA.107.042615},
  url = {https://link.aps.org/doi/10.1103/PhysRevA.107.042615}
}

@article{havlivcek2019supervised,
  title={Supervised learning with quantum-enhanced feature spaces},
  author={Havl{\'\i}{\v{c}}ek, Vojt{\v{e}}ch and C{\'o}rcoles, Antonio D and Temme, Kristan and Harrow, Aram W and Kandala, Abhinav and Chow, Jerry M and Gambetta, Jay M},
  journal={Nature},
  volume={567},
  number={7747},
  pages={209--212},
  year={2019},
  publisher={Nature Publishing Group}
}

@misc{Hangleiter_Advantage,
      title={Has quantum advantage been achieved?}, 
      author={Dominik Hangleiter},
      year={2026},
      eprint={2603.099015},
      archivePrefix={arXiv},
      primaryClass={quant-ph},
      url={https://arxiv.org/abs/2603.099015}, 
}

@misc{Fraud_Kyriienko,
      title={Unsupervised quantum machine learning for fraud detection}, 
      author={Oleksandr Kyriienko and Einar B. Magnusson},
      year={2022},
      eprint={2208.01203},
      archivePrefix={arXiv},
      primaryClass={quant-ph},
      url={https://arxiv.org/abs/2208.01203}, 
}

@misc{MiyabeFinance,
      title={Quantum Multiple Kernel Learning in Financial Classification Tasks}, 
      author={Shungo Miyabe and Brian Quanz and Noriaki Shimada and Abhijit Mitra and Takahiro Yamamoto and Vladimir Rastunkov and Dimitris Alevras  and Mekena Metcalf and Daniel J.M. King and Mohammad Mamouei and Matthew D. Jackson and Martin Brown and Philip Intallura and Jae-Eun Park},
      year={2023},
      eprint={2312.00260},
      archivePrefix={arXiv},
      primaryClass={quant-ph},
      url={https://arxiv.org/abs/2312.00260}, 
}

@article{VidalLC,
  title = {Class of Quantum Many-Body States That Can Be Efficiently Simulated},
  author = {Vidal, G.},
  journal = {Phys. Rev. Lett.},
  volume = {101},
  issue = {11},
  pages = {110501},
  numpages = {4},
  year = {2008},
  month = {Sep},
  publisher = {American Physical Society},
  doi = {10.1103/PhysRevLett.101.110501},
  url = {https://link.aps.org/doi/10.1103/PhysRevLett.101.110501}
}

@article{MarkovShi,
   title={Simulating Quantum Computation by Contracting Tensor Networks},
   volume={38},
   ISSN={1095-7111},
   url={http://dx.doi.org/10.1137/050644756},
   DOI={10.1137/050644756},
   number={3},
   journal={SIAM Journal on Computing},
   publisher={Society for Industrial & Applied Mathematics (SIAM)},
   author={Markov, Igor L. and Shi, Yaoyun},
   year={2008},
   month=Jan, pages={963–981} }

@misc{UtroToxicity,
      title={Enhanced Prediction of CAR T-Cell Cytotoxicity with Quantum-Kernel Methods}, 
      author={Filippo Utro and Meltem Tolunay and Kahn Rhrissorrakrai and Tanvi P. Gujarati and Jie Shi and Sara Capponi and Mirko Amico and Nate Earnest-Noble and Laxmi Parida},
      year={2025},
      eprint={2507.22710},
      archivePrefix={arXiv},
      primaryClass={quant-ph},
      url={
https://doi.org/10.48550/arXiv.2507.22710
}, 
}

@misc{TimeSeriesPolche,
      title={Hybrid Quantum-Classical Machine Learning Algorithms for Multi-Output Time-Series Forecasting at Utility Scale}, 
      author={Mackenson Polché and Varun Puram and Aditi Lal and Weronika Golletz and Joan Étude Arrow and Vardaan Sahgal and Kumar Ghosh, Giorgio Cortiana and Corey O'Meara},
      year={2026},
      eprint={2605.24252},
      archivePrefix={arXiv},
      primaryClass={quant-ph},
      url={
https://doi.org/10.48550/arXiv.2605.24252
}, 
}

@misc{SakhnenkoArtificial,
      title={Is data-efficient learning feasible with quantum models?}, 
      author={Alona Sakhnenko and Christian B. Mendl and Jeanette M. Lorenz},
      year={2026},
      eprint={2508.19437},
      archivePrefix={arXiv},
      primaryClass={quant-ph},
      url={https://arxiv.org/abs/2508.19437}, 
}

@article{Egginger_Hyperparam,
   title={A hyperparameter study for quantum kernel methods},
   volume={6},
   url={https://link.springer.com/article/10.1007/s42484-024-00172-1#citeas},
   DOI={10.1007/s42484-024-00172-1},
   number={4},
   journal={Quantum Machine Intelligence},
   publisher={Springer Nature},
   author={Sebastian Egginger and Alona Sakhnenko and Jeanette Miriam Lorenz},
   year={2024},
    }

@article{IQM_Utility,
   title={Evidence for the utility of quantum computing before fault tolerance},
   url={https://www.nature.com/articles/s41586-023-06096-3#citeas},
   DOI={10.1038/s41586-023-06096-3},
   volume={618},
   journal={Nature},
   pages = {500-505},
   publisher={Springer Nature},
   author={Youngseok Kim and Andrew Eddins and Sajant Anand and Ken Xuan Wei and Ewout van den Berg and Sami Rosenblatt and Hasan Nayfeh and Yantao Wu and Michael Zaletel and Kristan Temme and Abhinav Kandala},
   year={2023},
    }

@article{singh2025resource,
  title={A resource efficient quantum kernel},
  author={Singh, Utkarsh and Laprade, Jean-Fr{\'e}d{\'e}ric and Goldberg, Aaron Z and Heshami, Khabat},
  journal={arXiv preprint arXiv:2507.03689},
  year={2025}
}

@article{
canatar2023bandwidth,
title={Bandwidth Enables Generalization in Quantum Kernel Models},
author={Abdulkadir Canatar and Evan Peters and Cengiz Pehlevan and Stefan M. Wild and Ruslan Shaydulin},
journal={Transactions on Machine Learning Research},
issn={2835-8856},
year={2023},
url={https://openreview.net/forum?id=A1N2qp4yAq},
note={}
}

@article{peters2021machine,
  title={Machine learning of high dimensional data on a noisy quantum processor},
  author={Peters, Evan and Caldeira, Jo{\~a}o and Ho, Alan and Leichenauer, Stefan and Mohseni, Masoud and Neven, Hartmut and Spentzouris, Panagiotis and Strain, Doug and Perdue, Gabriel N},
  journal={npj Quantum Information},
  volume={7},
  number={1},
  pages={161},
  year={2021},
  publisher={Nature Publishing Group UK London}
}

@article{slattery2023numerical,
  title={Numerical evidence against advantage with quantum fidelity kernels on classical data},
  author={Slattery, Lucas and Shaydulin, Ruslan and Chakrabarti, Shouvanik and Pistoia, Marco and Khairy, Sami and Wild, Stefan M},
  journal={Physical Review A},
  volume={107},
  number={6},
  pages={062417},
  year={2023},
  publisher={APS}
}

@article{cerezo2024,
  title={Does provable absence of barren plateaus imply classical simulability? Or, why we need to rethink variational quantum computing},
  author={Marco Cerezo and Martin Larocca and Diego García-Martín and N. L. Diaz and Paolo Braccia and Enrico Fontana and Manuel S. Rudolph and Pablo Bermejo and Aroosa Ijaz and Supanut Thanasilp and Eric R. Anschuetz and Zoë Holmes},
  journal={Nature Communications, 16, 7907},
  year={2025}
}

@misc{StanfordHAI2026,
  author      = {{Stanford Institute for Human-Centered Artificial Intelligence}},
  title       = {Artificial Intelligence Index Report 2026},
  institution = {Stanford University},
  year        = {2026},
  address     = {Stanford, CA},
  url         = {https://hai.stanford.edu/assets/files/ai_index_report_2026.pdf}
}

@article{Abanin2025,
  author  = {Abanin, Dmitry A. and Acharya, Rajeev
             and Aghababaie-Beni, Laleh and others},
  title   = {Observation of constructive interference at the edge
             of quantum ergodicity},
  journal = {Nature},
  volume  = {646},
  number  = {8086},
  pages   = {825--830},
  year    = {2025},
  month   = oct,
  doi     = {10.1038/s41586-025-09526-6},
  url     = {https://doi.org/10.1038/s41586-025-09526-6}
}

@INPROCEEDINGS{MLTrends,
  author={Sevilla, Jaime and Heim, Lennart and Ho, Anson and Besiroglu, Tamay and Hobbhahn, Marius and Villalobos, Pablo},
  booktitle={2022 International Joint Conference on Neural Networks (IJCNN)}, 
  title={Compute Trends Across Three Eras of Machine Learning}, 
  year={2022},
  volume={},
  number={},
  pages={1-8},
  doi={10.1109/IJCNN55064.2022.9891914}}

@article{schnabel2025quantum,
  title={Quantum kernel methods under scrutiny: a benchmarking study},
  author={Schnabel, Jan and Roth, Marco},
  journal={Quantum Machine Intelligence},
  volume={7},
  number={1},
  pages={58},
  year={2025},
  publisher={Springer}
}

@article{muser2024provable,
  title={Provable advantages of kernel-based quantum learners and quantum preprocessing based on Grover's algorithm},
  author={Muser, Till and Zapusek, Elias and Belis, Vasilis and Reiter, Florentin},
  journal={Physical Review A},
  volume={110},
  number={3},
  pages={032434},
  year={2024},
  publisher={APS}
}

@article{liu2021rigorous,
  title={A rigorous and robust quantum speed-up in supervised machine learning},
  author={Liu, Yunchao and Arunachalam, Srinivasan and Temme, Kristan},
  journal={Nature physics},
  volume={17},
  number={9},
  pages={1013--1017},
  year={2021},
  publisher={Nature Publishing Group UK London}
}

@article{hubregtsen2022training,
  title={Training quantum embedding kernels on near-term quantum computers},
  author={Hubregtsen, Thomas and Wierichs, David and Gil-Fuster, Elies and Derks, Peter-Jan HS and Faehrmann, Paul K and Meyer, Johannes Jakob},
  journal={Physical Review A},
  volume={106},
  number={4},
  pages={042431},
  year={2022},
  publisher={APS}
}

@article{ThanasilpConcentration,
  title={Exponential concentration in quantum kernel methods},
  author={Supanut Thanasilp and Samson Wang and M. Cerezo and Zoë Holmes},
  journal={Nature Communications},
  volume={15},
  year={2024},
  publisher={Springer Nature}
}

@article{HEPKernel1,
  title = {Application of quantum machine learning using the quantum kernel algorithm on high energy physics analysis at the LHC},
  author = {Wu, Sau Lan and Sun, Shaojun and Guan, Wen and Zhou, Chen and Chan, Jay and Cheng, Chi Lung and Pham, Tuan and Qian, Yan and Wang, Alex Zeng and Zhang, Rui and Livny, Miron and Glick, Jennifer and Barkoutsos, Panagiotis Kl. and Woerner, Stefan and Tavernelli, Ivano and Carminati, Federico and Di Meglio, Alberto and Li, Andy C. Y. and Lykken, Joseph and Spentzouris, Panagiotis and Chen, Samuel Yen-Chi and Yoo, Shinjae and Wei, Tzu-Chieh},
  journal = {Phys. Rev. Res.},
  volume = {3},
  issue = {3},
  pages = {033221},
  numpages = {9},
  year = {2021},
  month = {Sep},
  publisher = {American Physical Society},
  doi = {10.1103/PhysRevResearch.3.033221},
  url = {https://link.aps.org/doi/10.1103/PhysRevResearch.3.033221}
}

@article{HEPKernel2,
  author  = {Belis, Vasilis and Wo{\'z}niak, Kinga Anna
             and Puljak, Ema and Barkoutsos, Panagiotis
             and Dissertori, G{\"u}nther and Grossi, Michele
             and Pierini, Maurizio and Reiter, Florentin
             and Tavernelli, Ivano and Vallecorsa, Sofia},
  title   = {Quantum anomaly detection in the latent space
             of proton collision events at the {LHC}},
  journal = {Communications Physics},
  volume  = {7},
  number  = {1},
  pages   = {334},
  year    = {2024},
  doi     = {10.1038/s42005-024-01811-6},
  url     = {https://doi.org/10.1038/s42005-024-01811-6}
}

@article{AQACYB,
  title={Pareto-optimal quantum kernel selection for unsupervised anomaly detection on real malware beaconing data},
  author={Boaz Micah and Nadia Milazzo and Maissa Beji and Borja Aizpurua and Llorenç Espinosa-Portalés and Esteban Payares and Ghada Ben Slama  and Luc Andrea  and Michel Kurek and Thomas Cope and Olivier Salomon},
  journal={Nature Communications},
  volume={15},
  year={2024},
  publisher={Springer Nature}
}

@article{Bremner2025,
  title = {Instantaneous Quantum Polynomial-Time Sampling and Verifiable Quantum Advantage: Stabilizer Scheme and Classical Security},
  author = {Bremner, Michael J. and Cheng, Bin and Ji, Zhengfeng},
  journal = {PRX Quantum},
  volume = {6},
  issue = {2},
  pages = {020315},
  numpages = {31},
  year = {2025},
  month = {Apr},
  publisher = {American Physical Society},
  doi = {10.1103/PRXQuantum.6.020315},
  url = {https://link.aps.org/doi/10.1103/PRXQuantum.6.020315}
}

@article{shepherd2009temporally,
  title={Temporally unstructured quantum computation},
  author={Shepherd, David and Bremner, Michael J.},
  journal={Proceedings of the Royal Society A: Mathematical, Physical and Engineering Sciences},
  volume={465},
  number={2106},
  pages={1413--1439},
  year={2009},
  publisher={The Royal Society},
  eprint={0809.0847},
  archivePrefix={arXiv}
}

@article{Shaw2024,
  author  = {Shaw, Adam L. and Chen, Zhuo
             and Choi, Joonhee and Mark, Daniel K.
             and Scholl, Pascal and Finkelstein, Ran
             and Elben, Andreas and Choi, Soonwon
             and Endres, Manuel},
  title   = {Benchmarking highly entangled states on a
             60-atom analogue quantum simulator},
  journal = {Nature},
  year    = {2024},
  volume  = {628},
  number  = {8006},
  pages   = {71--77},
  doi     = {10.1038/s41586-024-07173-x},
  url     = {https://doi.org/10.1038/s41586-024-07173-x}
}

@article{Andersen2025,
  author  = {Andersen, T. I. and Astrakhantsev, N.
             and Karamlou, A. H. and Berndtsson, J.
             and Motruk, J. and Szasz, A.
             and Gross, J. A. and Schuckert, A.
             and Westerhout, T. and Zhang, Y.
             and others},
  title   = {Thermalization and criticality on an
             analogue--digital quantum simulator},
  journal = {Nature},
  year    = {2025},
  volume  = {638},
  number  = {8049},
  pages   = {79--85},
  doi     = {10.1038/s41586-024-08460-3},
  url     = {https://doi.org/10.1038/s41586-024-08460-3}
}

@misc{Preskill_2012,
    author = {Preskill, J.},
    title = {{Quantum computing and the entanglement frontier}},
    year = {2012},
    note = {Rapporteur Talk at the 25th Solvay Conference on Physics, Brussels},
    doi = {https://doi.org/10.1142/8674},
}

@article{Preskill_2018,
    author = {Preskill, J.},
    title = {{Quantum computing in the NISQ era and beyond}},
    publisher = {{Verein zur F{\"{o}}rderung des Open Access Publizierens in den Quantenwissenschaften}},
    journal = {{Quantum}},
    volume = {2},
    pages = {79},
    year = {2018},
    doi = {10.22331/q-2018-08-06-79},
}

@article{
Many_Body_Problems,
author = {Hsin-Yuan Huang  and Richard Kueng  and Giacomo Torlai  and Victor V. Albert  and John Preskill },
title = {Provably efficient machine learning for quantum many-body problems},
journal = {Science},
volume = {377},
number = {6613},
pages = {eabk3333},
year = {2022},
doi = {10.1126/science.abk3333},
URL = {https://www.science.org/doi/abs/10.1126/science.abk3333},
eprint = {https://www.science.org/doi/pdf/10.1126/science.abk3333},
}

@article{wang2021towards,
  title={Towards understanding the power of quantum kernels in the NISQ era},
  author={Wang, Xinbiao and Du, Yuxuan and Luo, Yong and Tao, Dacheng},
  journal={Quantum},
  volume={5},
  pages={531},
  year={2021},
  publisher={Verein zur F{\"o}rderung des Open Access Publizierens in den Quantenwissenschaften}
}

@article{Phases_Support,
  title = {Quantum kernels to learn the phases of quantum matter},
  author = {Sancho-Lorente, Teresa and Rom\'an-Roche, Juan and Zueco, David},
  journal = {Phys. Rev. A},
  volume = {105},
  issue = {4},
  pages = {042432},
  numpages = {9},
  year = {2022},
  month = {Apr},
  publisher = {American Physical Society},
  doi = {10.1103/PhysRevA.105.042432},
  url = {https://link.aps.org/doi/10.1103/PhysRevA.105.042432}
}

@article{schuld2021supervised,
  title={Supervised quantum machine learning models are kernel methods},
  author={Schuld, Maria},
  journal={arXiv preprint arXiv:2101.11020},
  year={2021}
}

@article{Beverland22,
  title={Supervised quantum machine learning models are kernel methods},
  author={Michael E. Beverland and Prakash Murali and Matthias Troyer and Krysta M. Svore and Torsten Hoefler and Vadym Kliuchnikov and Guang Hao Low and Mathias Soeken and Aarthi Sundaram and Alexander Vaschillo},
  journal={arXiv preprint arXiv:2211.07629},
  year={2022}
}

@article{Gidney2021howtofactorbit,
  doi = {10.22331/q-2021-04-15-433},
  url = {https://doi.org/10.22331/q-2021-04-15-433},
  title = {How to factor 2048 bit {RSA} integers in 8 hours using 20 million noisy qubits},
  author = {Gidney, Craig and Eker{\aa{}}, Martin},
  journal = {{Quantum}},
  issn = {2521-327X},
  publisher = {{Verein zur F{\"{o}}rderung des Open Access Publizierens in den Quantenwissenschaften}},
  volume = {5},
  pages = {433},
  month = apr,
  year = {2021}
}

@article{PhysRevLett.127.180501,
  title = {Strong Quantum Computational Advantage Using a Superconducting Quantum Processor},
  author = {Wu, Yulin and Bao, Wan-Su and Cao, Sirui and Chen, Fusheng and Chen, Ming-Cheng and Chen, Xiawei and Chung, Tung-Hsun and Deng, Hui and Du, Yajie and Fan, Daojin and Gong, Ming and Guo, Cheng and Guo, Chu and Guo, Shaojun and Han, Lianchen and Hong, Linyin and Huang, He-Liang and Huo, Yong-Heng and Li, Liping and Li, Na and Li, Shaowei and Li, Yuan and Liang, Futian and Lin, Chun and Lin, Jin and Qian, Haoran and Qiao, Dan and Rong, Hao and Su, Hong and Sun, Lihua and Wang, Liangyuan and Wang, Shiyu and Wu, Dachao and Xu, Yu and Yan, Kai and Yang, Weifeng and Yang, Yang and Ye, Yangsen and Yin, Jianghan and Ying, Chong and Yu, Jiale and Zha, Chen and Zhang, Cha and Zhang, Haibin and Zhang, Kaili and Zhang, Yiming and Zhao, Han and Zhao, Youwei and Zhou, Liang and Zhu, Qingling and Lu, Chao-Yang and Peng, Cheng-Zhi and Zhu, Xiaobo and Pan, Jian-Wei},
  journal = {Phys. Rev. Lett.},
  volume = {127},
  issue = {18},
  pages = {180501},
  numpages = {7},
  year = {2021},
  month = {Oct},
  publisher = {American Physical Society},
  doi = {10.1103/PhysRevLett.127.180501},
  url = {https://link.aps.org/doi/10.1103/PhysRevLett.127.180501}
}

@article{Morvan2024,
  author  = {Morvan, A. and Villalonga, B. and Mi, X.
             and Mandr{\`a}, S. and Bengtsson, A.
             and Klimov, P. V. and Chen, Z.
             and Hong, S. and Erickson, C. and et al.},
  title   = {Phase transitions in random circuit sampling},
  journal = {Nature},
  year    = {2024},
  volume  = {634},
  number  = {8033},
  pages   = {328--333},
  doi     = {10.1038/s41586-024-07998-6},
  url     = {https://doi.org/10.1038/s41586-024-07998-6}
}

@article{chowdhury2024enhancing,
  title={Enhancing quantum utility: Simulating large-scale quantum spin chains on superconducting quantum computers},
  author={Chowdhury, Talal Ahmed and Yu, Kwangmin and Shamim, Mahmud Ashraf and Kabir, ML and Sufian, Raza Sabbir},
  journal={Physical Review Research},
  volume={6},
  number={3},
  pages={033107},
  year={2024},
  publisher={APS}
}

@article{lee2026digital,
  title={Digital quantum simulation of spin transport},
  author={Lee, Yi-Ting and Pokharel, Bibek and Cohn, Jeffrey and Schleife, Andr{\'e} and Banerjee, Arnab},
  journal={Physical review letters},
  volume={136},
  number={5},
  pages={050603},
  year={2026},
  publisher={APS}
}

@article{VidalMPS,
  title = {Efficient Simulation of One-Dimensional Quantum Many-Body Systems},
  author = {Vidal, Guifr{\'e}},
  journal = {Phys. Rev. Lett.},
  volume = {93},
  issue = {4},
  pages = {040502},
  numpages = {4},
  year = {2004},
  month = {Jul},
  publisher = {American Physical Society},
  doi = {10.1103/PhysRevLett.93.040502},
  url = {https://link.aps.org/doi/10.1103/PhysRevLett.93.040502}
}

@article{PGMPS,
author = {Perez-Garcia, D. and Verstraete, F. and Wolf, M. M. and Cirac, J. I.},
title = {Matrix product state representations},
year = {2007},
issue_date = {July 2007},
publisher = {Rinton Press, Incorporated},
address = {Paramus, NJ},
volume = {7},
number = {5},
issn = {1533-7146},
journal = {Quantum Info. Comput.},
month = jul,
pages = {401–430},
numpages = {30}
}

@article{yang2026quantum,
  title={Quantum simulation of many-body dynamics with noise-robust Trotter decomposition based on symmetric structures},
  author={Yang, Bo and Negishi, Naoki},
  journal={AVS Quantum Science},
  volume={8},
  number={1},
  year={2026},
  publisher={AIP Publishing}
}

@article{aktar2026structure,
  title={Structure-aware transformers for learning near-optimal trotter orderings with system-size generalization in 1d heisenberg hamiltonians},
  author={Aktar, Shamminuj and Tate, Reuben and Eidenbenz, Stephan},
  journal={arXiv preprint arXiv:2604.27171},
  year={2026}
}

@article{osborne2006,
  title = {Efficient Approximation of the Dynamics of One-Dimensional Quantum Spin Systems},
  author = {Osborne, Tobias J.},
  journal = {Phys. Rev. Lett.},
  volume = {97},
  issue = {15},
  pages = {157202},
  numpages = {4},
  year = {2006},
  month = {Oct},
  publisher = {American Physical Society},
  doi = {10.1103/PhysRevLett.97.157202},
  url = {https://link.aps.org/doi/10.1103/PhysRevLett.97.157202}
}

@article{vidal2007,
  title = {Classical Simulation of Infinite-Size Quantum Lattice Systems in One Spatial Dimension},
  author = {Vidal, G.},
  journal = {Phys. Rev. Lett.},
  volume = {98},
  issue = {7},
  pages = {070201},
  numpages = {4},
  year = {2007},
  month = {Feb},
  publisher = {American Physical Society},
  doi = {10.1103/PhysRevLett.98.070201},
  url = {https://link.aps.org/doi/10.1103/PhysRevLett.98.070201}
}

@article{haegeman2011,
  title = {Time-Dependent Variational Principle for Quantum Lattices},
  author = {Haegeman, Jutho and Cirac, J. Ignacio and Osborne, Tobias J. and Pi\ifmmode \check{z}\else \v{z}\fi{}orn, Iztok and Verschelde, Henri and Verstraete, Frank},
  journal = {Phys. Rev. Lett.},
  volume = {107},
  issue = {7},
  pages = {070601},
  numpages = {5},
  year = {2011},
  month = {Aug},
  publisher = {American Physical Society},
  doi = {10.1103/PhysRevLett.107.070601},
  url = {https://link.aps.org/doi/10.1103/PhysRevLett.107.070601}
}

@article{peng2022,
  title = {Quantum time dynamics employing the Yang-Baxter equation for circuit compression},
  author = {Peng, Bo and Gulania, Sahil and Alexeev, Yuri and Govind, Niranjan},
  journal = {Phys. Rev. A},
  volume = {106},
  issue = {1},
  pages = {012412},
  numpages = {12},
  year = {2022},
  month = {Jul},
  publisher = {American Physical Society},
  doi = {10.1103/PhysRevA.106.012412},
  url = {https://link.aps.org/doi/10.1103/PhysRevA.106.012412}
}

@article{wild2023,
  title = {Classical Simulation of Short-Time Quantum Dynamics},
  author = {Wild, Dominik S. and Alhambra, \'Alvaro M.},
  journal = {PRX Quantum},
  volume = {4},
  issue = {2},
  pages = {020340},
  numpages = {22},
  year = {2023},
  month = {Jun},
  publisher = {American Physical Society},
  doi = {10.1103/PRXQuantum.4.020340},
  url = {https://link.aps.org/doi/10.1103/PRXQuantum.4.020340}
}

@article{paeckel2019,
title = {Time-evolution methods for matrix-product states},
journal = {Annals of Physics},
volume = {411},
pages = {167998},
year = {2019},
issn = {0003-4916},
doi = {https://doi.org/10.1016/j.aop.2019.167998},
url = {https://www.sciencedirect.com/science/article/pii/S0003491619302532},
author = {Sebastian Paeckel and Thomas Köhler and Andreas Swoboda and Salvatore R. Manmana and Ulrich Schollwöck and Claudius Hubig}
}

@article{verresen2018,
  title = {Quantum dynamics of the square-lattice Heisenberg model},
  author = {Verresen, Ruben and Pollmann, Frank and Moessner, Roderich},
  journal = {Phys. Rev. B},
  volume = {98},
  issue = {15},
  pages = {155102},
  numpages = {13},
  year = {2018},
  month = {Oct},
  publisher = {American Physical Society},
  doi = {10.1103/PhysRevB.98.155102},
  url = {https://link.aps.org/doi/10.1103/PhysRevB.98.155102}
}

@article{marxer2023long,
  title={Long-distance transmon coupler with cz-gate fidelity above 99.8\%},
  author={Marxer, Fabian and Veps{\"a}l{\"a}inen, Antti and Jolin, Shan W and Tuorila, Jani and Landra, Alessandro and Ockeloen-Korppi, Caspar and Liu, Wei and Ahonen, Olli and Auer, Adrian and Belzane, Lucien and others},
  journal={PRX Quantum},
  volume={4},
  number={1},
  pages={010314},
  year={2023},
  publisher={APS}
}

@article{li2017efficient,
  title={Efficient variational quantum simulator incorporating active error minimization},
  author={Li, Ying and Benjamin, Simon C},
  journal={Physical Review X},
  volume={7},
  number={2},
  pages={021050},
  year={2017},
  publisher={APS}
}

@article{temme2017error,
  title={Error mitigation for short-depth quantum circuits},
  author={Temme, Kristan and Bravyi, Sergey and Gambetta, Jay M},
  journal={Physical review letters},
  volume={119},
  number={18},
  pages={180509},
  year={2017},
  publisher={APS}
}

@inproceedings{giurgica2020digital,
  title={Digital zero noise extrapolation for quantum error mitigation},
  author={Giurgica-Tiron, Tudor and Hindy, Yousef and LaRose, Ryan and Mari, Andrea and Zeng, William J},
  booktitle={2020 IEEE international conference on quantum computing and engineering (QCE)},
  pages={306--316},
  year={2020},
  organization={IEEE}
}

@article{mohammadipour2025direct,
  title={Direct analysis of zero-noise extrapolation: Polynomial methods, error bounds, and simultaneous physical-algorithmic error mitigation},
  author={Mohammadipour, Pegah and Li, Xiantao},
  journal={Quantum},
  volume={9},
  pages={1909},
  year={2025},
  publisher={Verein zur F{\"o}rderung des Open Access Publizierens in den Quantenwissenschaften}
}

@article{nation2021scalable,
  title={Scalable mitigation of measurement errors on quantum computers},
  author={Nation, Paul D and Kang, Hwajung and Sundaresan, Neereja and Gambetta, Jay M},
  journal={PRX Quantum},
  volume={2},
  number={4},
  pages={040326},
  year={2021},
  publisher={APS}
}

@article{maciejewski2020mitigation,
  title={Mitigation of readout noise in near-term quantum devices by classical post-processing based on detector tomography},
  author={Maciejewski, Filip B and Zimbor{\'a}s, Zolt{\'a}n and Oszmaniec, Micha{\l}},
  journal={Quantum},
  volume={4},
  pages={257},
  year={2020},
  publisher={Verein zur F{\"o}rderung des Open Access Publizierens in den Quantenwissenschaften}
}

@article{zhang2014protected,
  title={Protected quantum computing: interleaving gate operations with dynamical decoupling sequences},
  author={Zhang, Jingfu and Souza, Alexandre M and Brandao, Frederico Dias and Suter, Dieter},
  journal={Physical review letters},
  volume={112},
  number={5},
  pages={050502},
  year={2014},
  publisher={APS}
}

@article{pokharel2018demonstration,
  title={Demonstration of fidelity improvement using dynamical decoupling with superconducting qubits},
  author={Pokharel, Bibek and Anand, Namit and Fortman, Benjamin and Lidar, Daniel A},
  journal={Physical review letters},
  volume={121},
  number={22},
  pages={220502},
  year={2018},
  publisher={APS}
}

@misc{iqm_qubit_selector,
  author  = {{IQM Finland Oy}},
  title   = {{IQM Qubit Selector}},
  year    = {2026},
  url     = {https://pypi.org/project/iqm-qubit-selector/},
  note    = {Software package}
}

@article{schuld2019quantum,
  title={Quantum machine learning in feature Hilbert spaces},
  author={Schuld, Maria and Killoran, Nathan},
  journal={Physical review letters},
  volume={122},
  number={4},
  pages={040504},
  year={2019},
  publisher={APS}
}

@inproceedings{scholkopf2001generalized,
  title={A generalized representer theorem},
  author={Sch{\"o}lkopf, Bernhard and Herbrich, Ralf and Smola, Alex J},
  booktitle={International conference on computational learning theory},
  pages={416--426},
  year={2001},
  organization={Springer}
}

@book{scholkopf2001learning,
  title={Learning with kernels: support vector machines, regularization, optimization, and beyond},
  author={Sch{\"o}lkopf, Bernhard and Smola, Alexander J},
  year={2001},
  publisher={the MIT Press}
}

@article{JMLR:v13:cortes12a,
  author  = {Corinna Cortes and Mehryar Mohri and Afshin Rostamizadeh},
  title   = {Algorithms for Learning Kernels Based on Centered Alignment},
  journal = {Journal of Machine Learning Research},
  year    = {2012},
  volume  = {13},
  number  = {28},
  pages   = {795-828},
  url     = {http://jmlr.org/papers/v13/cortes12a.html}
}

@article{nadeau1999inference,
  title={Inference for the generalization error},
  author={Nadeau, Claude and Bengio, Yoshua},
  journal={Advances in neural information processing systems},
  volume={12},
  year={1999}
}

@article{bandyopadhyay2026provable,
  title={Provable learning separation for predicting time-evolution of quantum many-body systems},
  author={Bandyopadhyay, Rahul and Molteni, Riccardo and Eisert, Jens and Dunjko, Vedran and Jerbi, Sofiene},
  journal={arXiv preprint arXiv:2607.06472},
  year={2026}
}

@article{huang2020predicting,
  title={Predicting many properties of a quantum system from very few measurements},
  author={Huang, Hsin-Yuan and Kueng, Richard and Preskill, John},
  journal={Nature Physics},
  volume={16},
  number={10},
  pages={1050--1057},
  year={2020},
  publisher={Nature Publishing Group}
}

@misc{zenodo_data,
  author    = {Payares, Esteban and Milazzo, Nadia and Ben Slama, Ghada and Cope, Thomas},
  title     = {Data for ``Engineered Learning Problems on Noisy Quantum Processors''},
  year      = {2026},
  publisher = {Zenodo},
  version   = {1.0},
  doi       = {10.5281/zenodo.24394890},
  url       = {https://doi.org/10.5281/zenodo.24394890}
}

@misc{github_repo,
  author       = {Payares, Esteban and Milazzo, Nadia and Ben Slama, Ghada and Cope, Thomas},
  title        = {pqk-engineered-learning},
  year         = {2026},
  howpublished = {\url{https://github.com/PCesteban/pqk-engineered-learning}}
}

\clearpage
\onecolumngrid
\appendix

\section{Hyperparameters and Classical adversaries}\label{app:adversaries}
We benchmark against eleven classical baselines, each trained on the identical principal components and labels provided to the quantum kernels (Table~\ref{tab:hyperparams}).

\begin{table}[h]
    \centering
    \caption{Hyperparameter search spaces. All prediction models select their hyperparameters by 5-fold cross-validation on the $100$ training
  points (the test-tuned variant of Table~\ref{tab:gap_test} selects them on test accuracy instead). $d$ and $\sigma_x^2$: number and pooled
  variance of the input features; $n_f$ and $\sigma_f^2$: the same for the projected features. $\mathcal{C} = \{0.006, 0.015, 0.03\} \cup \{2^{-4},
  \dots, 2^{10}\}$ ($18$ values), $\mathcal{M} = \{0.25, 0.5, 1, 2, 3, 4, 5, 20\}$.}
    \label{tab:hyperparams}
    \resizebox{\textwidth}{!}{%
    \begin{tabular}{lll}
      \toprule
      Model & Hyperparameter & Search space \\
      \midrule
      Label construction & $\gamma_Q$ (referee PQK) & $\mathcal{M}/(n_f\sigma_f^2)$, maximin (Sec.~\ref{subsec:hyperparams}) \\
                         & $\gamma_C$ (adversary RBF) & $\{2^{-2}, \dots, 2^{6}\}/(d\sigma_x^2)$ \\
                         & $\lambda$ & $100$ log-spaced values in $[10^{-9}, 1]$, $g_{\mathrm{tra}} < 0.045$ \\
      \midrule
      PQK (SVM) & $\gamma_0$ & $\mathcal{M}$, Eq.~\eqref{eq:bandwidth} \\
                & $C$ & $\mathcal{C}$ \\
      \midrule
      SVM, RBF (labelling $\gamma$) & $\gamma$; $C$ & fixed at $\gamma_C^*$; $\mathcal{C}$ \\
      SVM, RBF (tuned) & $\gamma$; $C$ & $\mathcal{M}/(d\sigma_x^2)$; $\mathcal{C}$ \\
      SVM, Laplacian & $\gamma$; $C$ & $\mathcal{M}/(d\sigma_x^2)$; $\mathcal{C}$ \\
      SVM, linear, polynomial (deg.\ 2, 3), sigmoid, cosine & $C$ & $\mathcal{C}$ \\
      \midrule
      Gradient boosting & iterations; learning rate & $\{100, 300\}$; $\{0.05, 0.1\}$ \\
      Random forest & trees; max.\ depth & $\{200, 500\}$; $\{\text{none}, 5\}$ \\
      MLP (standardized inputs) & hidden layers; $L_2$ & $\{(64), (128, 64)\}$; $\{10^{-3}, 10^{-1}\}$ \\
      \bottomrule
    \end{tabular}%
    }
\end{table}

All hyperparameters are selected via 5-fold cross-validation over the training set, mirroring the PQK tuning protocol. As an upper-bound robustness check, Table~\ref{tab:gap_test} additionally reports margins relative to a test-tuned variant of each classical baseline whose hyperparameters are optimized directly on test accuracy, favoring the classical models. We observe similar behaviour to \cite{huang2021power,Gupta_2025}, where all classical methods perform broadly similar: with cross-validation the strongest is gradient boosting, while with test-set tuning the classical kernels are among the best performers.

\section{Unmitigated results: additional material}\label{app:unmit}
As circuit depth increases, the performance gap narrows considerably less at $n = 30$ and $45$ qubits than at $n = 15$ qubits (Table~\ref{tab:gap_test}). The challenger lags behind its referee by a comparable margin across all system sizes: for $n = 15$, $30$, and $45$ respectively, the referee–challenger performance discrepancy widens from $0.02$, $0.04$ and $0.04$ at $T = 3$ to $0.16$, $0.13$ and $0.12$ at $T = 10$. The different behaviour at larger sizes therefore comes mostly from the other two contributions to the gap: the referee's accuracy on its own labels and the accuracy of the best classical model.

\begin{figure}[ht]
    \centering
    \subfloat[$n{=}30$\label{fig:trotter_30q}]{%
      \includegraphics[width=0.49\textwidth]{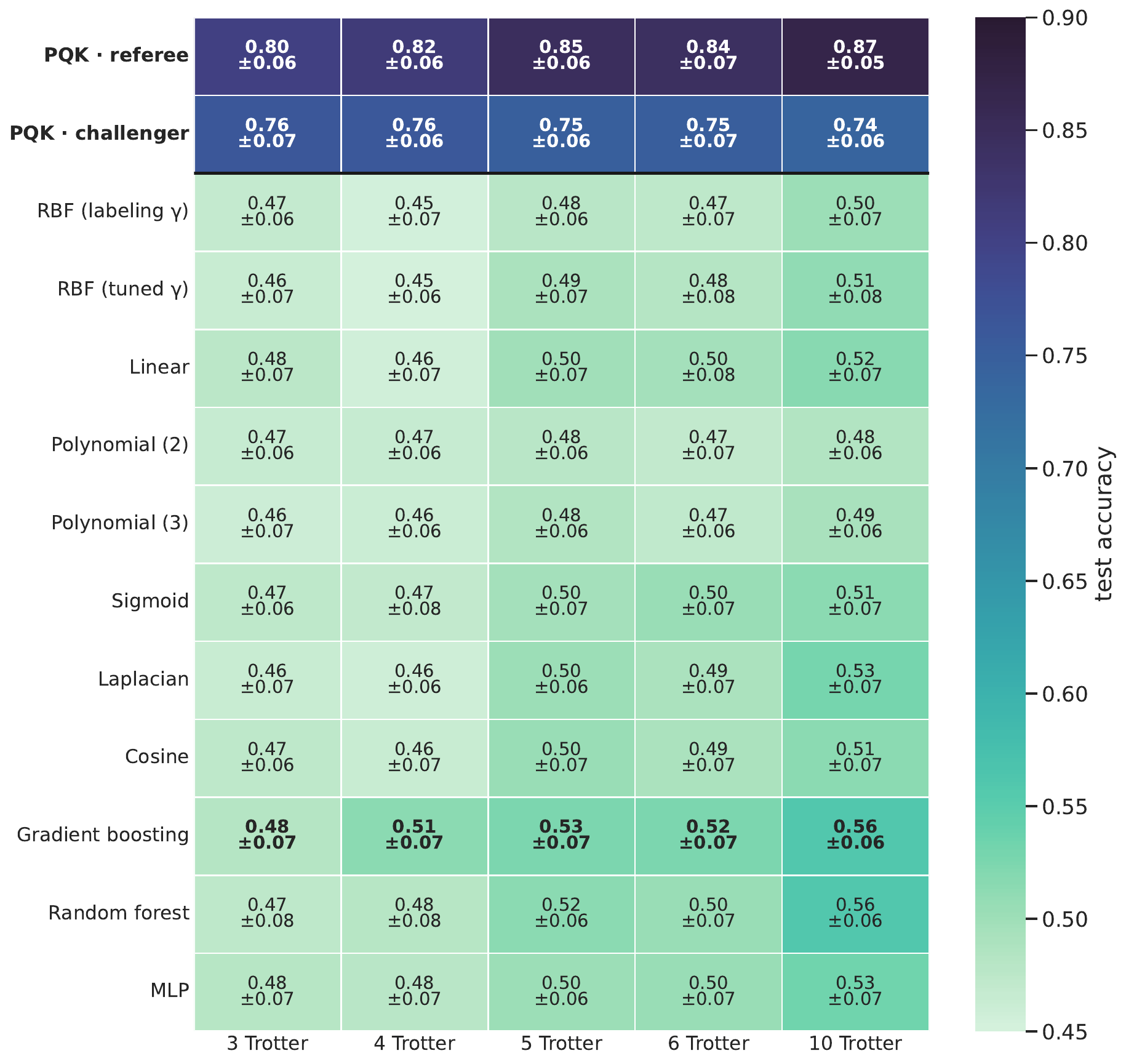}}\hfill
    \subfloat[$n{=}45$\label{fig:trotter_45q}]{%
      \includegraphics[width=0.49\textwidth]{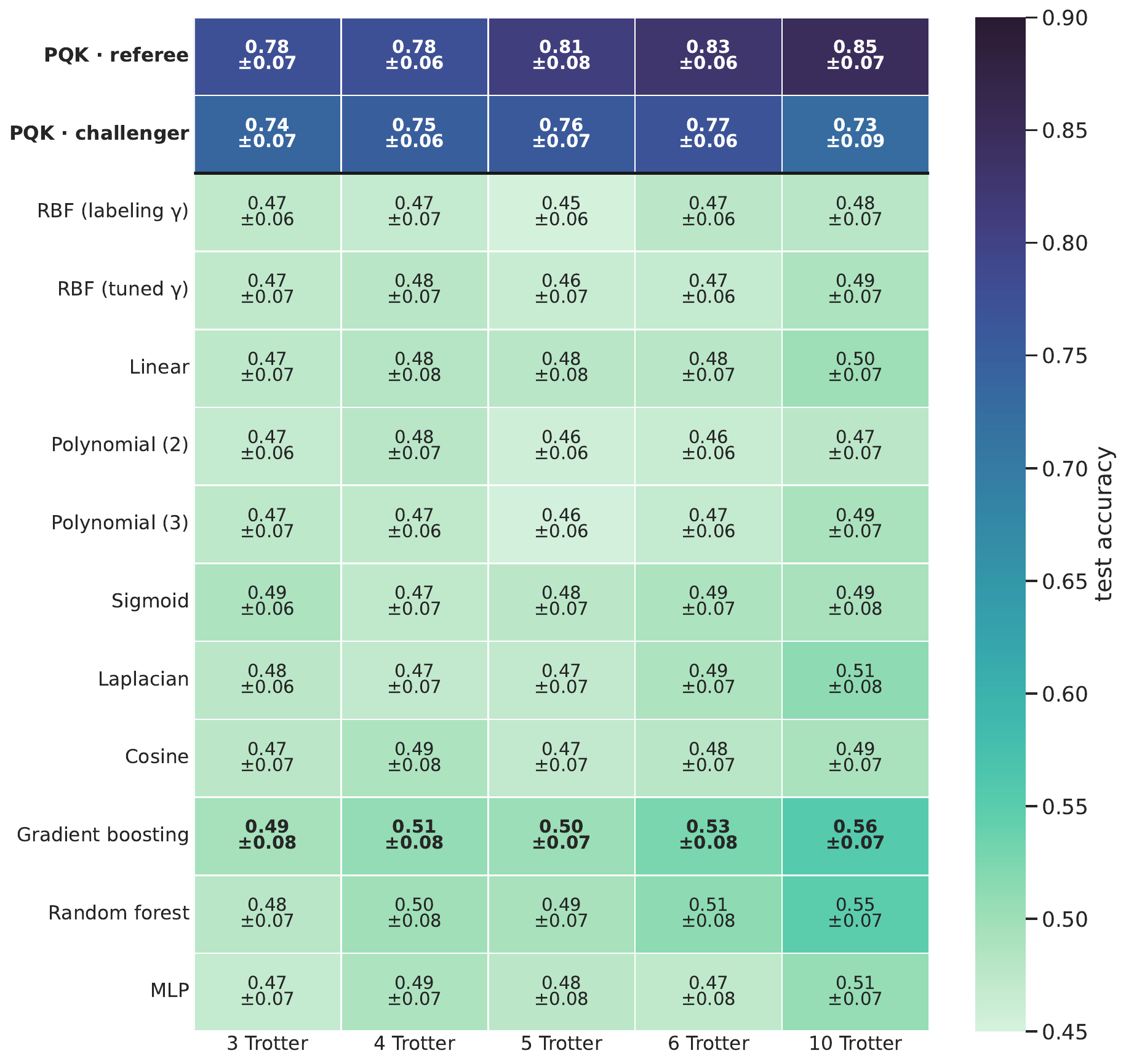}}
    \caption{As Fig.~\ref{fig:trotter_15q} at $n{=}30$~(a) and $n{=}45$~(b); referee and challenger are two chains on the same chip. Unlike at $n{=}15$, the gap persists at every depth: the challenger still falls behind its referee as $T$ grows, but the referee's accuracy on its own labels rises and the classical models improve less, so the gap narrows only slightly.}
    \label{fig:trotter_30q45q}
\end{figure}

At $n = 30$ and $45$, the referee's accuracy on its own labels increases with depth, while the top classical model improves less between $T = 3$ and $T = 10$ (see Fig.~\ref{fig:trotter_30q45q}); the smaller narrowing of the gap at these sizes therefore comes in large part from the labels.

\begin{figure}[t]
  \centering
  \subfloat[referee (chain A)\label{fig:chains_30q_a}]{%
    \includegraphics[width=0.49\textwidth]{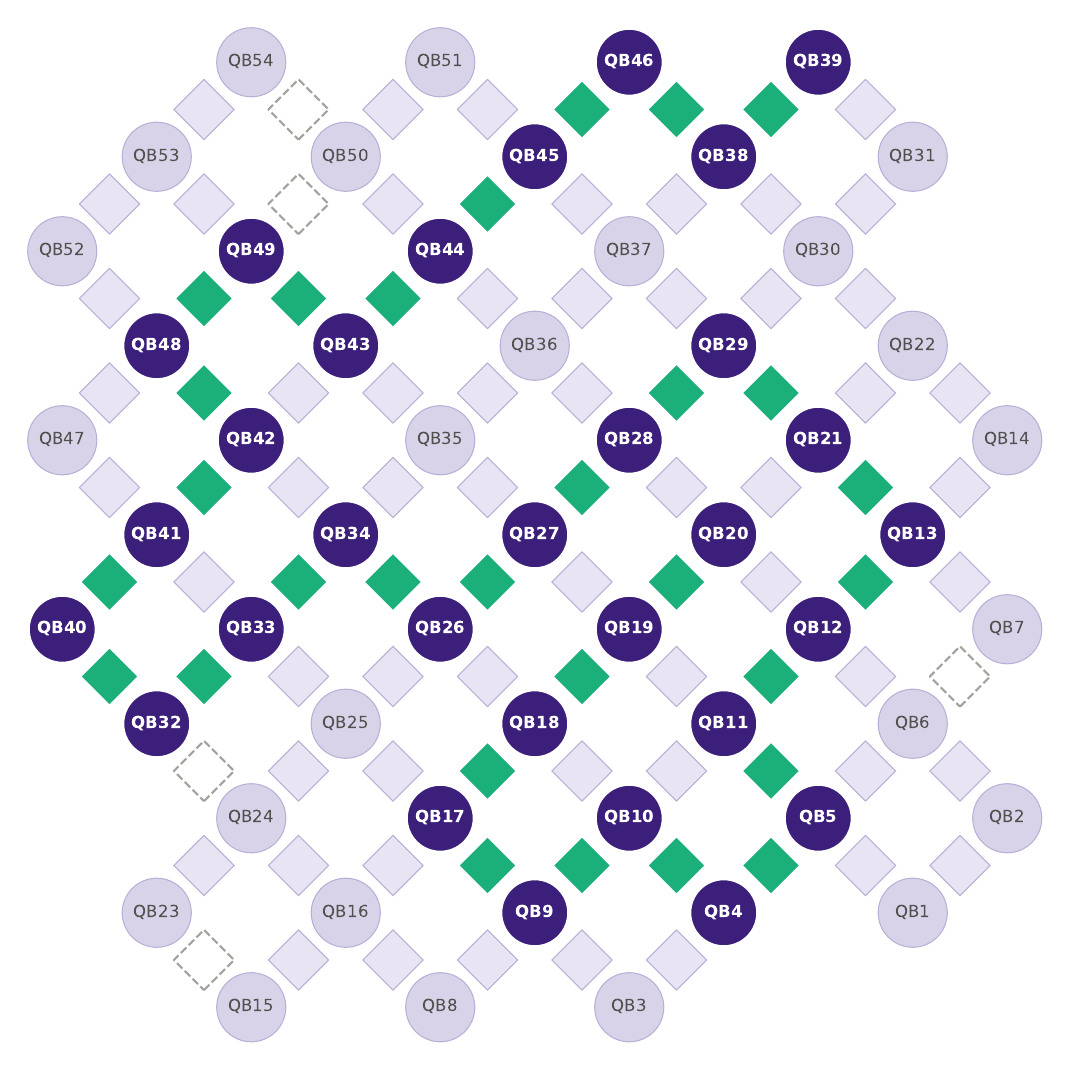}}\hfill
  \subfloat[challenger (chain B)\label{fig:chains_30q_b}]{%
    \includegraphics[width=0.49\textwidth]{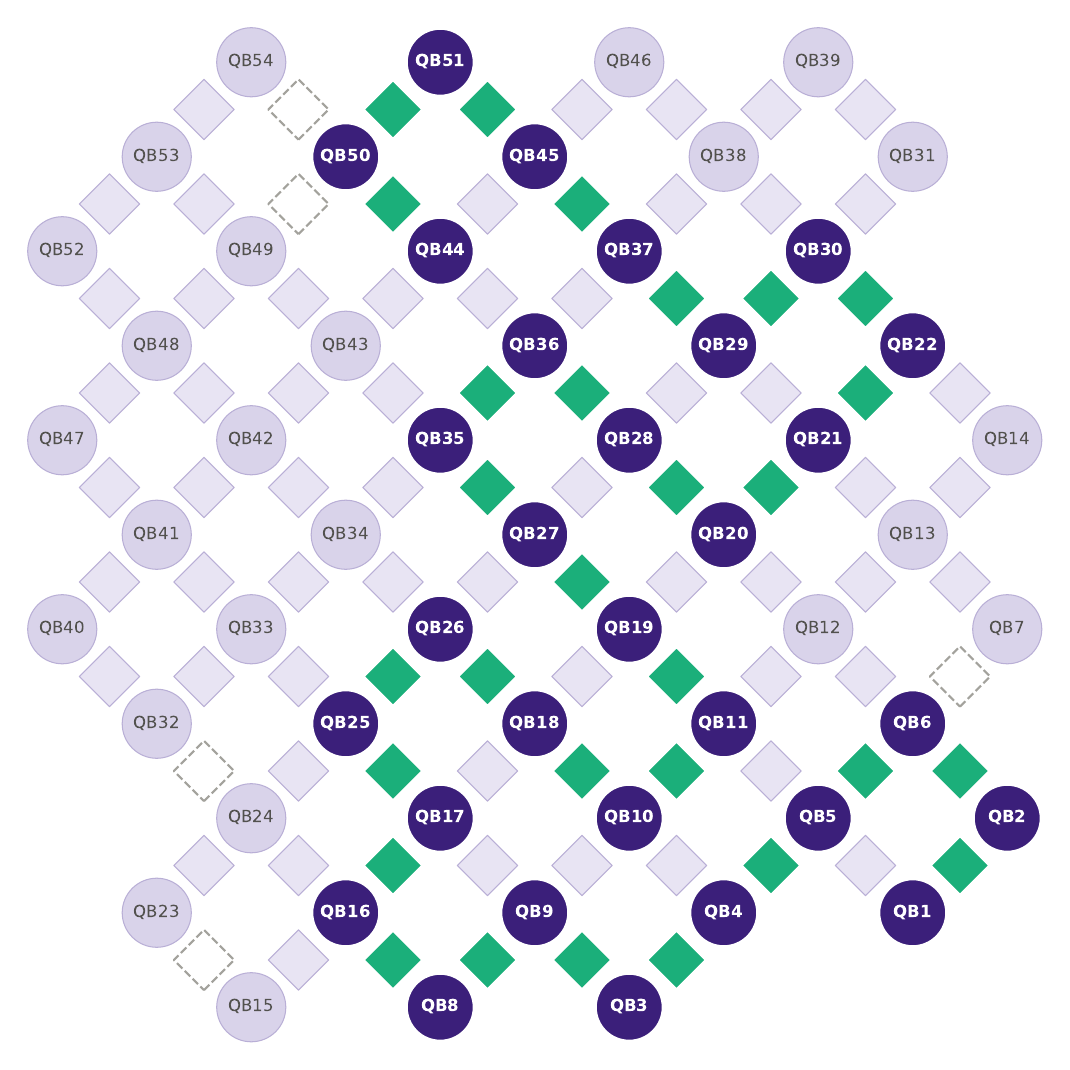}}
  \caption{The two $30$-qubit chains on \texttt{Emerald}, drawn as in Fig.~\ref{fig:emerald_chain}. They share $16$ of $30$ qubits and $1$ of
$29$ couplers.}
  \label{fig:chains_30q}
\end{figure}

\begin{figure}[t]
  \centering
  \subfloat[referee (chain A)\label{fig:chains_45q_a}]{%
    \includegraphics[width=0.49\textwidth]{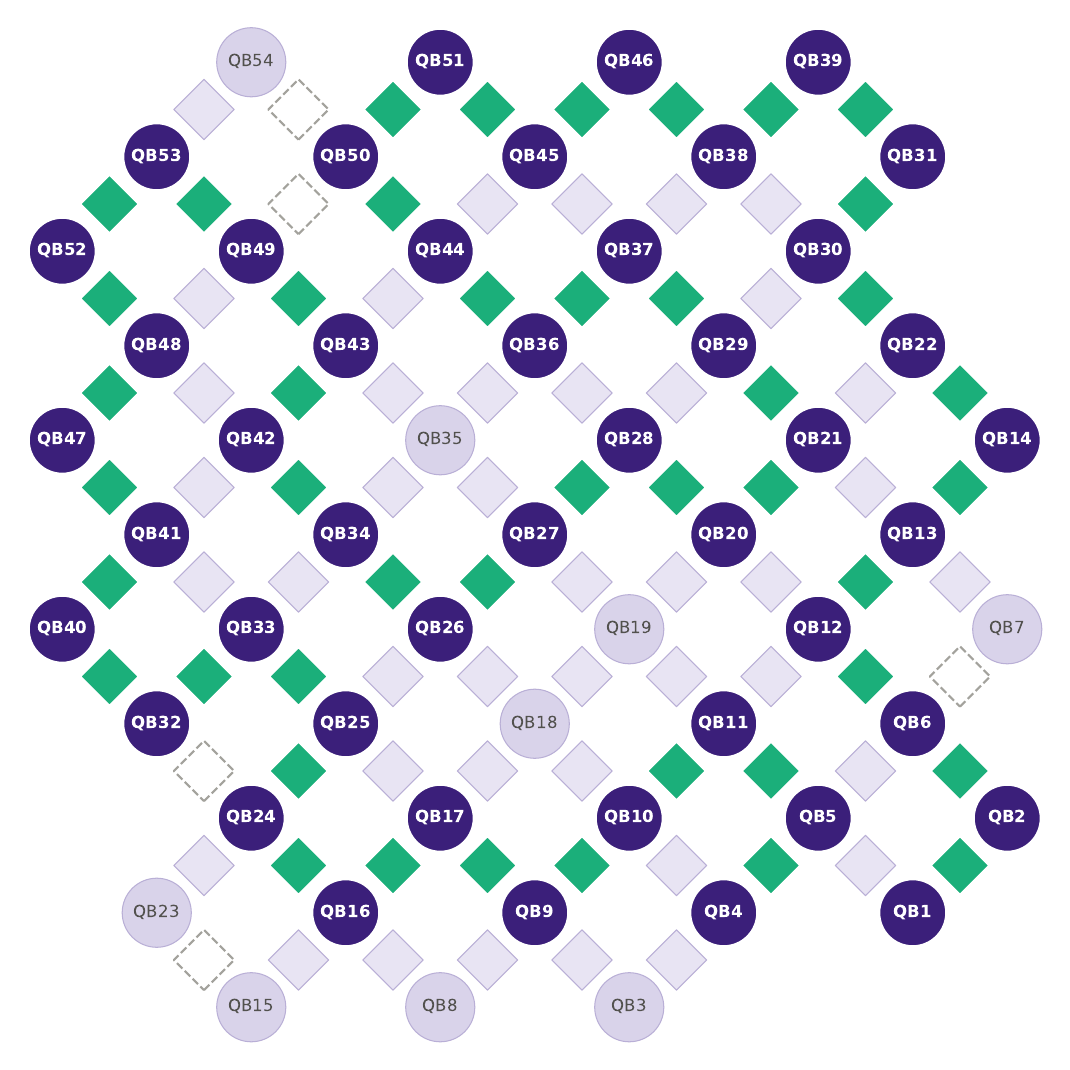}}\hfill
  \subfloat[challenger (chain B)\label{fig:chains_45q_b}]{%
    \includegraphics[width=0.49\textwidth]{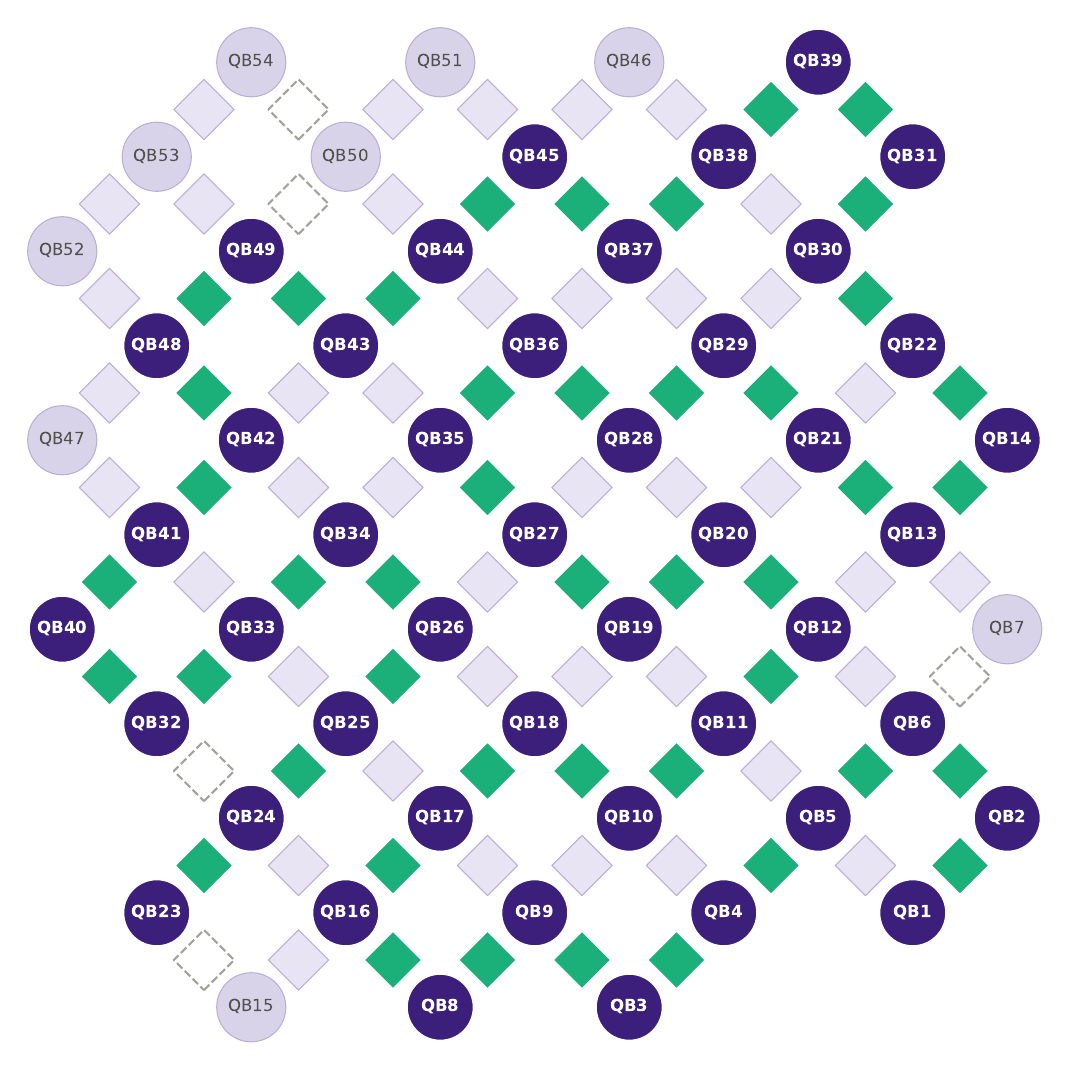}}
  \caption{As Fig.~\ref{fig:chains_30q} for the two $45$-qubit chains. They share $39$ of $45$ qubits and $18$ of $44$ couplers.}
  \label{fig:chains_45q}
\end{figure}

The somewhat larger referee–challenger discrepancy at $n = 15$ is consistent with the experimental setup: at $T = 10$, the cross-device challenger at $n = 15$ exhibits greater representational drift from its referee ($\mathrm{CKA} = 0.33$, compared to $0.54$ and $0.64$ for the same-chip pairs (Figs.~\ref{fig:chains_30q} and~\ref{fig:chains_45q}) at $n = 30$ and $45$, down from $\approx 0.88$ across all configurations at $T = 3$). This behaviour is consistent with what we observe on different devices, although here for different noise profiles on the same device given the larger number of qubits. Finally, when evaluating test-tuned classical models, the gap spans $+0.12$ to $+0.21$ across depths for both larger sizes, settling at $+0.12$ for $n = 30$ and $+0.14$ for $n = 45$ at $T = 10$.

\begin{table}[h]
  \centering
  \caption{CKA to the ideal kernel and effective error $\varepsilon$ (Sec.~\ref{subsec:compare_ideal}) of the 15-qubit hardware kernels versus Trotter depth, on the 150 points shared by all ideal simulations, together with the gap of Fig.~\ref{fig:trotter_15q}. E: \texttt{Emerald}, G: \texttt{Garnet}.}
  \label{tab:depth_ideal}
  \begin{tabular}{lccccc}
    \toprule
    $T$ & 3 & 4 & 5 & 6 & 10 \\
    \midrule
    CKA to ideal, E    & $0.963$ & $0.946$ & $0.919$ & $0.868$ & $0.669$ \\
    CKA to ideal, G    & $0.928$ & $0.894$ & $0.857$ & $0.801$ & $0.593$ \\
    \midrule
    $\varepsilon$, E   & $0.210$ & $0.245$ & $0.302$ & $0.369$ & $0.604$ \\
    $\varepsilon$, G   & $0.285$ & $0.344$ & $0.415$ & $0.469$ & $0.758$ \\
    \midrule
    gap (CV)           & $+0.300$ & $+0.287$ & $+0.248$ & $+0.184$ & $+0.040$ \\
    \bottomrule
  \end{tabular}
\end{table}

\begin{figure}[ht]
    \centering
    \includegraphics[width=0.8\textwidth]{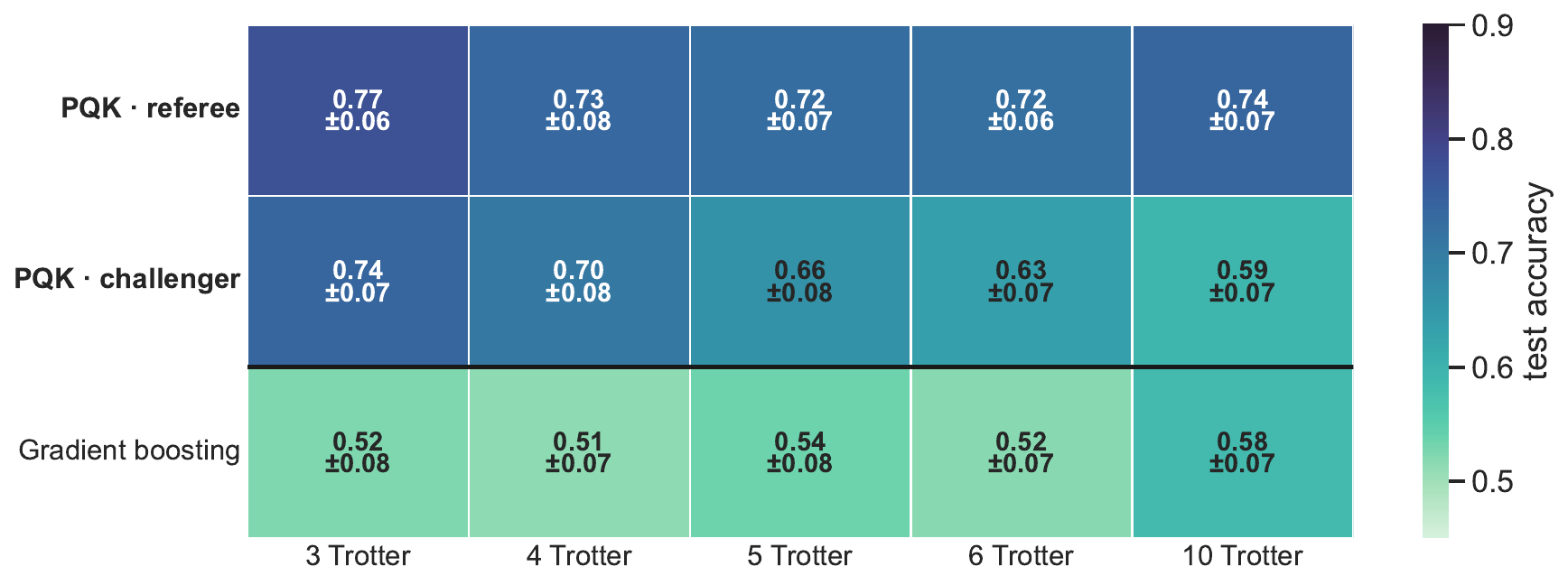}
    \caption{Test accuracy on new engineered labels constructed via Eq.~\eqref{eq:newlabels} at $n{=}15$ for $T{=}3$--$10$, mean $\pm$ s.d.\ (Sec.~\ref{subsec:compare_ideal}). Rows: the referee on its own labels, the challenger, and the best performing classical model among the ones reported in Appendix~\ref{app:adversaries}. This new label construction results in a decrease in quantum performance.} 
    \label{fig:newlabels}
\end{figure}

We also include here the results of the test accuracies for the experiment conducted on labels generated via the function introduced in Sec.~\ref{subsec:newlabels}, which can generalise to new test points, not originally included in the dataset of size $N$. We report here only the accuracies of the quantum kernels and the best classical model found (see Fig.~\ref{fig:newlabels}). The performance of the PQKs is lower on these new labels compared to the ones in Fig.~\ref{fig:trotter_15q}.

\section{Error mitigation: additional material}\label{app:em}
\begin{figure}[ht]
  \centering
  \includegraphics[width=0.8\textwidth]{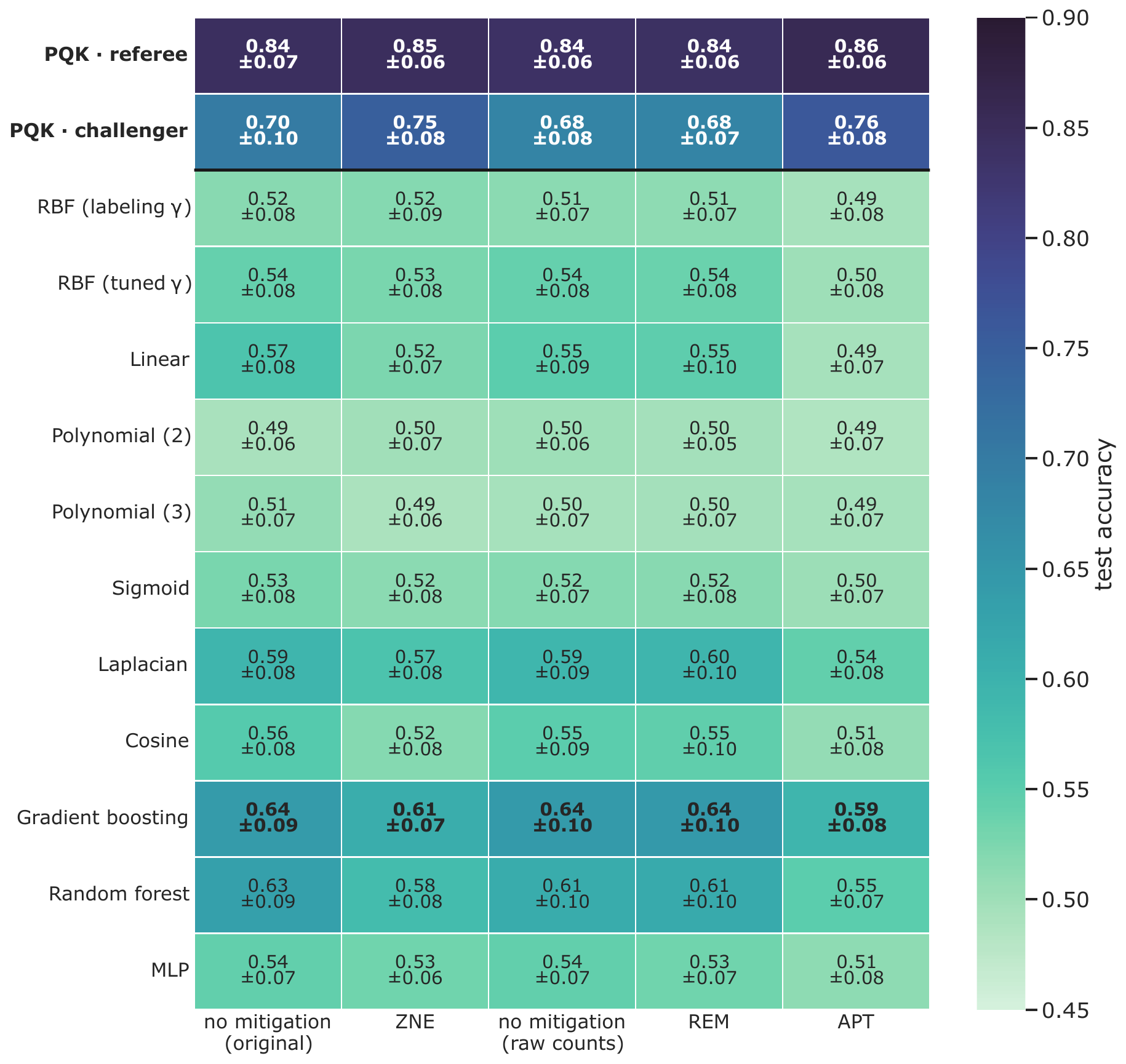}
  \caption{Mean test accuracy $\pm$ s.d.\ (Sec.~\ref{subsec:compare_ideal}) on engineered labels at $n{=}15$, $T{=}10$. Columns: no mitigation and ZNE (original run); no mitigation, REM and APT (raw counts). Rows: the referee on its own labels (\texttt{Emerald}), the challenger (\texttt{Garnet}), and the classical models.}
  \label{fig:em_heatmap}
\end{figure}

Figure~\ref{fig:em_heatmap} shows the test accuracy of every model in the error mitigation experiments of Sec.~\ref{subsubsec:em_results}: the referee on its own labels, the challenger, and all classical models, for each mitigation method and its baseline.

\subsection{Earlier configurations}\label{app:em_earlier}

The effect of ZNE on features and kernels reported in Sec.~\ref{subsubsec:em_results} is not specific to that configuration. Before the main experiment, which uses direct measurement at $n = 15$, we applied ZNE in six earlier configurations. These differ in measurement strategy (classical shadows~\cite{huang2020predicting}, which estimate the features from measurements in randomly chosen single-qubit Pauli bases, or direct measurement, which reads out all qubits in the $X$, $Y$ and $Z$ bases with three circuits per data point), number of data points ($N = 5$ to $150$), system size ($n = 5$, $15$ and $30$), depth ($T = 3$ and $5$) and device (Table~\ref{tab:em_appendix}). In all of them, ZNE moves $c$ toward $1$, while $\varepsilon$ changes by at most about $20\%$ and the distance $d_F$ to the ideal kernel by no more than $0.025$. The one exception is the $30$-qubit configuration, where $d_F$ nearly doubles ($0.163 \to 0.304$).

\begin{table}[ht]
    \centering
    \caption{ZNE versus no mitigation on the PQK in earlier configurations, compared with the
    noiseless kernel. Rows are $n{=}5$, $T{=}3$ on \texttt{Emerald} unless noted; $a\,/\,b$ is
    \texttt{Emerald} / \texttt{Garnet}. Columns as in Table~\ref{tab:em_main}. At $n{=}30$ the ideal
    kernel is an MPS simulation with bond dimension $128$.}
    \label{tab:em_appendix}
    \begin{tabular}{llcccc}
      \toprule
      Configuration & Mitig. & $c$ & $\lVert r\rVert$ & $\varepsilon$ & $d_F$ \\
      \midrule
      Shadows, $N{=}5$ & none & $0.75\pm0.03$ & $0.26\pm0.02$ & $0.35$ & $0.093\pm0.021$ \\
      (4 seeds)        & ZNE  & $0.95\pm0.04$ & $0.38\pm0.03$ & $0.40$ & $0.118\pm0.055$ \\
      \midrule
      Direct, $N{=}5$  & none & $0.79$ & $0.19$ & $0.24$ & $0.055$ \\
                       & ZNE  & $1.00$ & $0.25$ & $0.25$ & $0.057$ \\
      \midrule
      Direct, $N{=}50$ & none & $0.73$ & $0.25$ & $0.34$ & $0.157$ \\
                       & ZNE  & $0.93$ & $0.30$ & $0.32$ & $0.155$ \\
      \midrule
      Shadows, $n{=}15$, $N{=}150$ & none & $0.81$ & $0.23$ & $0.28$ & $0.106$ \\
                                   & ZNE  & $0.95$ & $0.29$ & $0.31$ & $0.124$ \\
      \midrule
      Direct, $n{=}15$, $T{=}5$, $N{=}150$ & none & $0.67\,/\,0.34$ & $0.31\,/\,0.27$ & $0.47\,/\,0.79$ & $0.168\,/\,0.205$ \\
                                           & ZNE  & $0.91\,/\,0.60$ & $0.44\,/\,0.39$ & $0.48\,/\,0.65$ & $0.171\,/\,0.199$ \\
      \midrule
      Shadows, $n{=}30$, $T{=}5$, $N{=}150$ & none & $0.55$ & $0.51$ & $0.92$ & $0.163$ \\
                                            & ZNE  & $0.76$ & $0.76$ & $1.00$ & $0.304$ \\
      \bottomrule
    \end{tabular}
\end{table}

\subsection{Ranking kernels by \texorpdfstring{$\varepsilon$}{ε} and \texorpdfstring{$c$}{c}}\label{app:em_rank}
Across the ten hardware kernels in Table~\ref{tab:em_main}, $\varepsilon$ predicts their distance to the ideal kernel more effectively than $c$ (Table~\ref{tab:em_rank}). While $c$ also provides a reasonable ordering, this is partly because the \texttt{Emerald} and \texttt{Garnet} kernels differ simultaneously in both metrics. Given the sample size of ten kernels evaluated across the two devices, these correlations should be viewed as descriptive of this specific benchmark rather than as a general rule.

\begin{table}[!ht]
    \centering
    \caption{Rank correlation (Spearman) between each quantity and the distance of the kernel to the ideal one, over the ten hardware kernels of Table~\ref{tab:em_main}.}
    \label{tab:em_rank}
    \begin{tabular}{lcc}
      \toprule
      & $d_F$ to ideal & CKA to ideal \\
      \midrule
      $\varepsilon$ & $+0.95$ & $-0.98$ \\
      $c$           & $-0.60$ & $+0.73$ \\
      \bottomrule
    \end{tabular}
\end{table}

\subsection{Geometry of the mitigation effect}\label{app:em_geometry}

We use the centred kernel alignment (CKA), introduced in \cite{JMLR:v13:cortes12a}. Its calculation begins by centering each feature vector by the mean feature vector, creating from the kernel matrix $K \in \mathbb{R}^{N \times N}$ a new kernel matrix:
\begin{equation}
\tilde{K} := (\mathbb{I} - \mathbf{1}/N) K (\mathbb{I} - \mathbf{1}/N),
\end{equation}
where $\mathbf{1}$ is the $N \times N$ matrix of all ones. The CKA is then defined as the cosine similarity of two of these matrices with respect to the Frobenius norm,
\begin{equation}\label{Eq:CKA}
\mathrm{CKA}(K, K') := \frac{\langle \tilde{K}, \tilde{K}' \rangle_F}{\sqrt{\langle \tilde{K}, \tilde{K} \rangle_F \langle \tilde{K}', \tilde{K}' \rangle_F}}.
\end{equation}

To decouple the overall magnitude of kernel deformation induced by mitigation from its alignment toward the ideal target, we treat the unmitigated, mitigated, and ideal kernels as points in metric space. We then compute the angle between the mitigation vector (unmitigated $\to$ mitigated) and the direction toward the ideal (unmitigated $\to$ ideal) using $d_F$ alongside the normalized  CKA distance, $d(K,K') = \sqrt{2 - 2\,\mathrm{CKA}(K,K')}$, evaluated over unit Frobenius-norm centred matrices.

Zero-noise extrapolation alters the kernels by a magnitude comparable to their initial distance from the ideal ($d_F = 0.194$ and $0.193$ from unmitigated to ZNE, compared to $0.233$ and $0.262$ toward the ideal). However, this shift points largely away from the ideal: the angular deviation is $55^\circ$ ($d_F$) and $54^\circ$ (CKA) for \texttt{Emerald}, and $76^\circ$ and $64^\circ$ for \texttt{Garnet}, yielding only a small reduction in distance to the ideal, or an actual increase for \texttt{Garnet} under $d_F$ (Table~\ref{tab:em_main}). In contrast, the APT effect, evaluated relative to the raw counts baseline, exhibits comparable or greater magnitude ($d_F = 0.228$ and $0.383$) while aligning substantially closer to the ideal direction ($37^\circ$ and $28^\circ$ for \texttt{Emerald}; $30^\circ$ and $41^\circ$ for \texttt{Garnet}). Consequently, it closes $28\%\text{--}52\%$ of the initial distance to the ideal across both kernels and metrics, compared to at most $15\%$ for ZNE.

\subsection{Where the ZNE gain comes from}\label{app:em_labels}
The ZNE referee and challenger kernels exhibit slightly greater similarity than their unmitigated counterparts: across the $600$ points, their CKA is $0.551$, compared to $0.496$ for the unmitigated pair and $0.524$ between the ZNE referee and unmitigated challenger. In terms of relative Frobenius error, however, all three pairs remain nearly equidistant ($0.244$, $0.248$, and $0.238$), indicating that the ZNE challenger is no closer to the ZNE referee than the unmitigated challenger. This observation aligns with their respective accuracies on the ZNE referee labels ($0.748$ vs.\ $0.740$): applying ZNE to the challenger yields minimal additional benefit.

The observed gain therefore originates primarily from the labels. This label effect (Sec.~\ref{subsubsec:em_results}) is not uniformly distributed across the ten subsamples. It ranges from $-0.09$ to $+0.22$ and does not follow how much ZNE alters the referee kernel, whose CKA relative to the unmitigated referee kernel lies between $0.61$ and $0.68$ across all subsamples.

Most of this label effect is concentrated in the two subsamples where the ZNE referee's maximin selection picks a different bandwidth multiplier from the unmitigated referee ($1.0$ instead of $0.5$): there, the unmitigated challenger gains $+0.180$ on average, compared to only $+0.005$ in the other eight, where both referees pick the same multiplier. To isolate the bandwidth selection from the kernel transformation, we regenerated the ZNE referee labels for these two subsamples with the bandwidth fixed to the unmitigated referee's choice, keeping all other parameters unchanged. Under this constraint, the label effect in these two subsamples drops from $+0.180$ to $+0.071$, reducing the mean effect across all ten subsamples from $+0.040$ to $+0.018$.

Roughly half of the overall label effect can therefore be attributed to the bandwidth chosen for labelling, while the remainder arises from the ZNE kernel itself, predominantly localized within a few subsamples. While the ZNE referee kernel is marginally closer to the challenger than the unmitigated referee in every subsample (CKA increase of $+0.01$ to $+0.04$), the magnitude of this shift only loosely follows the observed label effect.

\end{document}